\documentclass[aps,pra]{revtex4-1}
\usepackage{amsmath}
\usepackage[amsmath]{SIunits}

\begin{document}
\title{Effect of temperature anisotropy on various modes and instabilities
for a magnetized non-relativistic bi-Maxwellian plasma}
\author{M. F. Bashir$^{1,2}$ and G. Murtaza$^{1}$\\
  1. Salam Chair in
Physics, GC University Lahore, 54000, Pakistan 
2. Department of
Physics, GC University Lahore, 54000, Pakistan }
\begin{abstract}
  Using kinetic theory for homogeneous collisionless magnetized
  plasmas, we present an extended review of the plasma waves and
  instabilities and discuss the anisotropic response of generalized
  relativistic dielectric tensor and Onsager symmetry properties for
  arbitrary distribution functions. In general, we observe that for
  such plasmas only those electromagnetic modes whose magnetic-field
  perturbations are perpendicular to the ambient magnetic field, i.e.,
  $\mathbf{B}_{1}$ $\bot $ $\mathbf{B}_{0}$, are effected by the
  anisotropy. However, in oblique propagation all  modes do show such anisotropic effects. Considering the
  non-relativistic bi-Maxwellian distribution and studying the
  relevant components of the general dielectric tensor under
  appropriate conditions,  we derive the dispersion relations for
  various modes and instabilities. We show that only the
  electromagnetic R- and L- waves, those derived from them (i.e., the
  whistler mode, pure Alfv\'{e}n mode, firehose instability, and
  whistler instability), and the O-mode are affected by thermal
  anisotropies, since they satisfy the required condition
  $\mathbf{B}_{1}\bot \mathbf{B}_{0}$. By contrast, the
  perpendicularly propagating X-mode and the modes derived from it (
  the pure transverse X-mode and Bernstein mode) show no such effect. In
  general, we note that the thermal anisotropy modifies the parallel
  propagating modes via the parallel acoustic effect, while it
  modifies the perpendicular propagating modes via the Larmor-radius
  effect. In oblique propagation for kinetic Alfv\'{e}n waves, the
  thermal anisotropy affects the kinetic regime more than it affects
  the inertial regime. The generalized fast mode exhibits two distinct
  acoustic effects, one in the direction parallel to the ambient
  magnetic field and the other in the direction perpendicular to it. In
  the fast-mode instability, the magneto-sonic wave causes suppression
  of the firehose instability. We discuss all these propagation
  characteristics and present graphic illustrations. The
  threshold conditions for different instabilities are also obtained.
\newline
\newline
\newline
\footnotesize{Email: frazbashir@yahoo.com}

\end{abstract}

\maketitle

\section{Introduction}
\label{sec:3}
Extensive studies have been conducted over the years using a wide
variety of particle distributions to derive the general dielectric
tensor so as to study various modes and instabilities
\cite{chen}-\cite{verdon}. Brambilla \cite{brambilla} also noted that,
in a homogeneous collisionless plasma, the dielectric tensor always
satisfies the Onsager symmetry relations independently of the particular
form of the equilibrium distribution function.  Gaelzer et
al., \cite{gaelzer} presented a detailed derivation of the effective
longitudinal dielectric constant for plasmas in inhomogeneous magnetic
fields. Ziebell and Schneider \cite{zeibell} provided a general
expression for an effective dielectric tensor satisfying the Onsager
symmetry in an inhomogeneous plasma. Miyamoto \cite{miyamoto},
choosing an anisotropic streaming Maxwellian plasma, obtained the
general dielectric tensor and employed it to derive the dispersion
relation for different instabilities.

As is well known, the temperature is anisotropic in several
environments, such as the solar wind, solar corona, auroral
ionosphere, magnetosphere, astrophysical and space plasmas, and
plasmas produced in recent experimental developments. These naturally
occurring and laboratory plasmas generally have bi-Maxwellian or
nearly bi-Maxwellian velocity distributions. Noci et al.,~\cite{noci}~analyzed the
data from the outer solar corona to show that the velocity
distribution function becomes anisotropic beyond 1.8 solar
radii. Pagel et al., \cite{pagel} observed that the core population of
solar wind is well described by a bi-Maxwellian distribution function,
while the halo component is best modeled by a bi-Kappa distribution
function. Schunk and Watkins \cite{shunck} showed that the anisotropy
in the electron-temperature distribution in the ionosphere develops
above 2500\,\kilo\meter\ and increases with altitude
. Masood et al., \cite{masod} analysed that the electron-velocity
distribution in the Earth's magneto-sheath is thermally anisotropic. Cremaschini et al., \cite{Cremaschini} showed that the accretion-disk
plasmas around the compact and massive objects are characterized by
thermally anisotropic velocity distribution functions
. In the laboratory, the incident high-energy laser
wave can make the resulting plasma thermally anisotropic, because the
plasma is more intensely heated in the direction of the laser-wave
electric field \cite{Gillani}-\cite{Sid}.

The thermal anisotropy not only influences the propagation of
various modes, but also induces various types of electromagnetic
instabilities in a variety of plasmas including space and
astrophysical plasma, fusion plasmas ( both magnetic and inertial
confinement) as well as in the plasma created by highly intense free
electron x-ray laser pulses. In particular, the stability analysis of
\ whistler wave and Alfv\'{e}nic modes (pure Alfv\'{e}%
n wave, magnetosonic wave, kinetic Alfv\'{e}n waves, inertial
Alfv\'{e}n wave, etc.) have received special attention. Whistler waves
are electromagnetic waves in magnetized plasmas with frequencies below
the cyclotron frequency. Whistlers are naturally produced by lightning
discharges in thunderstorms. When produced near the north or south
pole, they can travel from one pole to another along the Earth's
magnetic lines of force through the ionosphere. As a result, lightning
flashes in the Southern Hemisphere can be observed in the Northern
Hemisphere \cite{chen}. Plasma-wave instruments on Voyager 2 have
detected whistler-mode emissions inside and outside the magnetosphere
of Saturn. Whistlers are also observed to propagate through
self-created ducts in magnetospheres \cite{xiao}.  Whistler modes are
used to induce radio-frequency plasma discharges, and to heat plasmas
in tokamaks\cite{pawan} and spheromaks\cite{stenzel}. The Weibel
instability and the Weibel instability in an ambient magnetic field
(i.e., the whistler instability), which have been known for several
decades \cite{weibel}, are of significant interest. Studies
considering different particle-velocity distributions have relied on
the Weibel \cite{weibel1}-\cite{Thaury} and whistler instabilities
\cite{sadia3}-\cite{lazar} to explain physical processes in different
plasma environments.

Alfv\'{e}n waves are believed to play major roles in certain
astrophysical processes in magnetized plasmas such as ones found in
the environments of stars and interstellar clouds
\cite{h.alfven}. Many phenomena in the solar atmosphere or
heliosphere, planetary and cometary magnetospheres, cometary tails,
Earth's ionosphere, etc., can also be regarded as manifestations of
linear or nonlinear Alfv\'{e}n waves. Due to their incompressibility
and low reflectivity in the solar atmosphere \cite{l.ofman},
Alfv\'{e}n waves have been invoked as the most promising wave
mechanism to explain the heating of Sun's outer atmosphere, or corona,
to millions of degrees and the acceleration of the solar wind to
hundreds of kilometers per second\cite{B.D}. Gekelman has studied
Alfv\'{e}n waves and their relationship to space observations in the
laboratory \cite{gekelman1}. One of the most important instabilities
that excite Alfv\'{e}n waves in a hot plasma is the so-called
fire-hose instability, driven by the plasma temperature
anisotropy. Since the thermal anisotropy is an intrinsic
characteristic of magnetized plasmas, especially in the cases of
collisionless plasmas such as astrophysical and space plasmas, the
fire-hose instability has many implications in these plasmas, and has
been discussed in the literature on the basis of the Vlasov kinetic
theory. The fire-hose instability in magnetized thermal plasmas has
been studied by a number of authors \cite{shilick
  fire1}-\cite{lazarfire1}. Kinetic Alfv\'{e}n waves (KAWs) are
produced due to charge separation when the perpendicular wavelength
becomes comparable to the ion gyroradius and show dispersive character
for oblique propagation \cite{hasegawabook}. Bashir et al., discussed
the effect of thermal anisotropy on the propagation characteristics of  KAWs in both the kinetic and the inerial regimes\cite{Bashir}. Various other
aspects of the KAWs have been studied by several authors
\cite{kawsinstability}-\cite{33}. Recent theoretical and experimental
advances focused on the wave propagation of KAWs in different regimes
have associated the morphology of the wave with the generation
mechanism proposed by Gekelman et al.,~\cite{34}.

Here, we present an extended review on plasma waves and instabilities
and discuss the anisotropic response of the general relativistic
dielectric tensor as well as the Onsager symmetric properties for a
homogeneous magnetized collisionless plasma for an arbitrary
distribution function.  Considering non-relativistic bi-Maxwellian
distributions, we simplify the analytical expressions for the
components of dielectric tensor in the limit $%
\left\vert \xi _{n\alpha }\right\vert \gg 1,$ where $\xi _{n\alpha
}=(\omega \,-n\Omega _{\alpha })/ (k_{\Vert }v_{t\Vert \alpha})$; this
allows easier switching to parallel or perpendicular propagation.  We
moreover obtain the dispersion relations for various modes and
instabilities, along with their graphical representation, from the
simplified components of the dielectric tensor under the appropriate
conditions.  For example, (i) for parallel propagation, we derive the
dispersion relations for the R- and L-waves, whistler wave, Alfv\'{e}n
wave, Langmuir wave, Alfv\'{e}n wave instability, whistler
instabilities, and Weibel instabilities, along with the conditions for
instability; (ii) for perpendicular propagation, the
general dispersion relations for the X-mode, O-mode and Bernstein mode
are derived; (iii) for oblique propagation, we derive the general
dispersion relations for the Kinetic Alfv\'{e}n waves (KAWs) in the
kinetic and inertial regimes, and for the fast mode. In addition, we
recover a number of special cases for the general fast-mode
dispersion relation under appropriate conditions. We also discuss the fast mode
instability for oblique propagation.

The plan of the paper is as follows. In Section~II, we present the
general relativistic dielectric tensor and then employ the
non-relativistic bi-Maxwellian distribution function to simplify the
general dispersion relation for various modes and instabilities under
appropriate conditions, along with their graphical representations. In
Section~III we briefly summarize and discuss the results.

\section{Mathematical Model}
\label{sec:5}
We start out with the relativistic Vlasov equation
\begin{equation}
\frac{\partial f_{\alpha }}{\partial t}+\mathbf{v}\cdot \frac{\partial
f_{\alpha }}{\partial \mathbf{x}}+q_{\alpha }(\mathbf{E}+\frac{\mathbf{%
v\times B}}{c})\cdot \frac{\partial f_{\alpha }}{\partial \mathbf{p}}=0
\label{1}
\end{equation}
where the relativistic momentum $\mathbf{p}$ and velocity $\mathbf{v}$ are
related by
\begin{eqnarray}
\mathbf{p} &=&\gamma m\mathbf{v}\mathrm{ \qquad }\gamma =\left( 1-\frac{v^{2}}{%
c^{2}}\right) ^{-\frac{1}{2}}=\sqrt{1+\frac{p^{2}}{m^{2}c^{2}}}  \nonumber \\
&&  \nonumber \\
\mathbf{v} &=&\frac{c\mathbf{p}}{\sqrt{m^{2}c^{2}+p^{2}}}  \label{2} \\
&&  \nonumber
\end{eqnarray}

Equation~(1), combined with the Maxwell Equations
\begin{eqnarray}
\mathbf{\nabla \times E} &=&-\frac{1}{c}\frac{\partial \mathbf{B}}{\partial t%
}\mathrm{ \quad }  \label{3} \\
&&  \nonumber \\
\mathrm{\quad }\mathbf{\nabla \times B} &=&\frac{1}{c}\frac{\partial \mathbf{E}%
}{\partial t}+\frac{4\pi }{c}\mathbf{J}  \label{4},
\end{eqnarray}
describe the dynamics of the plasma system.%
\[
\]

Linearizing Eqs.~(\ref{1}), (\ref{3}), and (\ref{4}), taking the Fourier-Laplace
transform, assuming the ambient magnetic field $\mathbf{B}_{0}$ along z-
direction and the wavenumber vector\ $\ \mathbf{k}$  in the x-z plane, we
obtain the perturbed distribution function 
\begin{eqnarray*}
f_{1\alpha } &=&\frac{q_{\alpha }}{\Omega _{\alpha }}\int\limits_{\pm \infty
}^{\phi }\exp \left[ \frac{1}{\Omega _{\alpha }}\left\{ \left( s+ik_{\Vert
}v_{\Vert }\right) (\phi -\phi ^{\prime })+ik_{\bot }v_{\bot }(\sin \phi
-\sin \phi ^{\prime })\right\} \right] \left( \mathrm{ }(\mathbf{E}_{1}+\frac{%
\mathbf{v}^{\prime }\mathbf{\times B}_{1}}{c})\cdot \frac{\partial
f_{0\alpha }}{\partial \mathbf{p}^{\prime }}\right) d\phi ^{\prime }
\end{eqnarray*}
and the dyadic equation
\begin{eqnarray}
\left( s^{2}+c^{2}k^{2}\right) \mathbf{E}_{1}\mathbf{-}c^{2}\mathbf{k}(%
\mathbf{k.E}_{1})+4\pi s\mathbf{J} &=&\mathbf{0}  \label{5} \\
&&  \nonumber
\end{eqnarray}%

Given the distribution function $f_{1\alpha }$, we can calculate the current density 
\[
\mathbf{J=}\overleftrightarrow{\mathbf{\sigma }}\mathbf{.E}%
_{1}=\sum\limits_{\alpha }q_{\alpha }n_{0\alpha }\int \mathbf{v}f_{1\alpha
}d^{3}p, 
\]%
where $\overleftrightarrow{\mathbf{\sigma }}$ is the conductivity tensor.

We can rewrite the dyadic Eq.~(\ref{5}) in terms of the
dielectric permittivity tensor $\overleftrightarrow{\varepsilon }$\cite{miyamoto} as 
\begin{equation}
\left[ \varepsilon _{ij}-N^{2}\left( \delta _{ij}-\frac{N_{i}N_{j}}{N^{2}}%
\right) \right] E_{j}=0,  \label{6}
\end{equation}
where 
\begin{eqnarray}
\epsilon _{ij} &\equiv &\delta _{ij}+\dfrac{4\pi \sigma _{ij}}{s}=\delta
_{ij}\,\mathbf{-}\sum\limits_{n=\mathrm{ }-\mathrm{ }\infty }^{\infty
}M_{ij}+L_{ij},\label{7}
\end{eqnarray}
with
\begin{eqnarray*}
M_{ij} &=&\frac{2\pi }{s}\sum_{\alpha }m_{\alpha }\omega _{p\alpha
}^{2}\int\limits_{-\infty }^{\infty }dp_{_{\Vert }}\int\limits_{0}^{\infty
}p_{\perp }dp_{\perp }\sum\limits_{n=\mathrm{ }-\mathrm{ }\infty }^{\infty }%
\frac{\,\chi _{1}}{s+ik_{_{\Vert }}v_{\Vert }+in\Omega _{\alpha }}\times \\
&&\times \left( 
\begin{array}{lll}
\begin{array}{l}
\begin{array}{l}
\begin{array}{l}
\dfrac{n^{2}}{z^{2}}v_{\perp }\left[ J_{n}(z)\right] ^{2}%
\end{array}%
\end{array}%
\end{array}
& 
\begin{array}{l}
\begin{array}{l}
\begin{array}{l}
\dfrac{in}{z}v_{\perp }J_{n}(z)J\,_{n}^{\prime }(z)%
\end{array}%
\end{array}%
\end{array}
& 
\begin{array}{l}
\begin{array}{l}
\begin{array}{l}
\quad \dfrac{n}{z}v_{\Vert }[J_{n}(z)]^{2}%
\end{array}%
\end{array}%
\end{array}
\\ 
\begin{array}{l}
\begin{array}{l}
\begin{array}{l}
-\dfrac{in}{z}v_{\perp }J_{n}(z)J\,_{n}^{\prime }(z)%
\end{array}%
\end{array}%
\end{array}
& 
\begin{array}{l}
\begin{array}{l}
\begin{array}{l}
v_{\perp }[J_{n}^{\prime }(z)]^{2}%
\end{array}%
\end{array}%
\end{array}
& 
\begin{array}{l}
\begin{array}{l}
\,-iv_{\Vert }J\,_{n}(z)J\,_{n}^{\prime }(z)%
\end{array}%
\end{array}

\\ 
\begin{array}{l}
\dfrac{n}{z}v_{\Vert }[J_{n}(z)]^{2}%
\end{array}
& 
\begin{array}{l}
iv_{\Vert }J\,_{n}(z)J\,_{n}^{\prime }(z)%
\end{array}
& 
\begin{array}{l}
\quad \quad \dfrac{v_{\Vert }^{2}}{v_{\bot }}[J_{n}(z)]^{2}%
\end{array}%
\end{array}%
\right),
\end{eqnarray*}
the only nonzero component of $L_{ij}$ being
\[
L_{zz}=\frac{2\pi }{s^{2}}\sum_{\alpha }m_{\alpha }\omega _{p\alpha
}^{2}\int\limits_{-\infty }^{\infty }dp_{_{\Vert }}\int\limits_{0}^{\infty
}p_{\perp }dp_{\perp }\left[ \frac{v_{\Vert }}{v_{\bot }}\left( v_{\Vert }%
\frac{\partial \,f_{0\alpha }}{\partial \,p_{\bot }}-v_{\bot }\frac{\partial
\,f_{0\alpha }}{\partial \,p_{\Vert }}\right) \right] 
\]
Here we have used the following properties of the Bessel functions:
\[
\sum\limits_{n=-\infty }^{\infty }n[J_{n}(z)]^{2}=0,\quad
\sum\limits_{n=-\infty }^{\infty }J_{n}(z)J_{n}^{^{\prime }}(z)=0,\quad
\sum\limits_{n=-\infty }^{\infty }[J_{n}(z)]^{2}=1,
\]
and defined
\[
\chi _{1\,}=\left\{ \frac{\partial \,f_{0\alpha }}{\partial \,p_{\bot }}+%
\frac{i\,k_{\Vert }}{s}\left( v_{\Vert }\frac{\partial \,f_{0\alpha }}{%
\partial \,p_{\bot }}-v_{\bot }\frac{\partial \,f_{0\alpha }}{\partial
\,p_{\Vert }}\right) \right\}.
\]

$N=\dfrac{ck}{\omega }$ is the total refractive index, whose
components are $N_{i,j}=ck_{i,\ j}/\omega$, while $\omega _{p\alpha
}^{2}=4\pi n_{o\alpha }e^{2}/\,m_{0\alpha }\,$\ is the non-relativistic
plasma frequency, and $\Omega _{\alpha }$ $[=\Omega _{0\alpha }/\gamma 
=q_{\alpha }B_{0}/(\gamma m_{0\alpha }c)$] is the relativistic
cyclotron frequency.\\

Equation~(\ref{7}) gives the general relativistic dielectric
tensor in a homogeneous magnetized plasma for an arbitrary equilibrium
distribution function $f_{0\alpha }$. From Eq.(\ref{7}), it is clear
that $\chi _{1}$ ( only for $k_{\Vert }\neq 0$) and $L_{zz}$
contain the anisotropic modification factor $\left( v_{\Vert
  }\dfrac{\partial \,f_{0\alpha }}{\partial \,p_{\bot }}-v_{\bot
  }\dfrac{\partial \,f_{0\alpha }%
  }{\partial \,p_{\Vert }}\right) $. This factor stems from the term
$\left( \mathbf{v}\times \mathbf{B}_{1}\right) \cdot \dfrac{\partial
  f_{0\alpha }}{%
  \partial \mathbf{p}}$ in the linearized Vlasov equation, which
vanishes for isotropic velocity distributions or under electrostatic
conditions, i.e., for $\mathbf{k}\Vert \mathbf{E}$. The same term
however survives for those electromagnetic modes in which the ambient
magnetic field is perpendicular to the perturbed magnetic field, i.e.,
for $\mathbf{B}_{1}$ $\bot $ $\mathbf{B}_{0}$, for any arbitrary
anisotropic distribution (from non-relativistic to relativistic
regimes). From the Maxwell Equations, it is clear that generation of
perturbed magnetic field depends upon the propagation direction and
the polarizations of electric field. As illustrated by Fig.~1 the R-
and L-waves and the waves derived from them for parallel propagation,
and the O-mode for perpendicular propagation satisfy the requirement
that the ambient magnetic field be perpendicular to the perturbed
magnetic field. The same is evident from Eq.(\ref{7}), since $\chi
_{1}$ exhibits the thermal anisotropy effect for $k_{\Vert }\neq
0$ only.  In the case of the O-mode (i.e., $\epsilon _{zz}-N_{\bot
}^{2}=0$), the anisotropy effect comes from the $L_{zz}$
component. For parallel propagation of electrostatic modes (i.e., for
$\epsilon _{zz}=0$), a contribution from the anisotropic effect is
found in $M_{ij}$ and another is found in $L_{ij}$, but the two
contributions cancel each other. The obliquely propagating electromagnetic modes always contain the
anisotropic modification.

 In the following section we will prove that for
a non-relativistic bi-Maxwellian distribution, only the modes whose 
perturbed magnetic field is perpendicular to the ambient field are
modified by the thermal anisotropy.

A remarkable property of the general dielectric tensor is
described by the Onsager symmetry relations,
\[
\epsilon _{xy}=-\epsilon _{yx},\ \ \epsilon _{xz}=\epsilon _{zx}\ \mathrm{and
\ }\epsilon _{yz}=-\epsilon _{zy} 
\]%
which is always satisfied by the hot-plasma dielectric tensor,
independently of the choice of equilibrium distribution
function $f_{0\alpha }.$

The above discussion on the anisotropic response of dielectric tensor and
Onsager symmetry relations are valid for arbitrary anisotropic
distributions and cover magnetized-plasma environments ranging from
the non-relativistic to the ultra-relativistic regimes.

\subsection{General Dielectric Tensor for Non-relativistic Bi-Maxwellian
Distribution}
\label{sec:4}
We shall now derive the general dielectric tensor for a
non-relativistic bi-Maxwellian plasma. The non-relativistic bi-Maxwellian
distribution function is 
\begin{equation}
f_{0\alpha }=\frac{1}{2\pi m_{\alpha }T_{\bot \alpha }\left( 2\pi m_{\alpha
}T_{\Vert \alpha }\right) ^{\frac{1}{2}}}\exp \left[ -\dfrac{p_{\perp }^{2}}{%
2m_{\alpha }T_{\perp \alpha }}-\frac{p_{\parallel }^{2}}{2m_{\alpha
}T_{\parallel \alpha }}\right]  \label{8}
\end{equation}

Once the integrations over $p_{\parallel}$ and $p_{\perp}$ are carried
out, with the help of the Onsager symmetry relations, the following general
dispersion relation is obtained \cite{Bashir}
\begin{equation}
\left\vert 
\begin{array}{lll}
\epsilon _{xx}-N_{\parallel }^{2} & \epsilon _{xy} & \epsilon
_{xz}+N_{\parallel }N_{\perp } \\ 
-\epsilon _{xy} & \epsilon _{yy}-N^{2} & \epsilon _{yz} \\ 
\epsilon _{xz}+N_{\parallel }N_{\perp } & -\epsilon _{yz} & \epsilon
_{zz}-N_{\perp }^{2}%
\end{array}%
\right\vert =0  \label{9}
\end{equation}
where
\begin{eqnarray}
&&  \nonumber \\
\epsilon _{xx} &=&1+\sum_{\alpha }\frac{\omega _{p\alpha }^{2}}{\omega ^{2}}%
\sum\limits_{n\,=\,-\,\infty }^{\infty }\dfrac{n^{2}}{\lambda _{\alpha }}%
\Gamma _{n}(\lambda _{\alpha })\left\{ \xi _{0\alpha }\,Z\left( \xi
_{n\alpha }\right) -\frac{Z^{\prime }\left( \xi _{n\alpha }\right) }{2}%
\,\left( \frac{T_{\bot \alpha }}{T_{\Vert \alpha }}-\,1\right) \right\} 
\label{10} \\
\varepsilon _{yy} &=&1+\sum_{\alpha }\frac{\omega _{p\alpha }^{2}}{\omega
^{2}}\sum\limits_{n\,=\,-\,\infty }^{\infty }\left( \frac{n^{2}\Gamma
_{n}(\lambda _{\alpha })}{\lambda _{\alpha }}-2\lambda _{\alpha }\,\Gamma
_{n}^{\prime }(\lambda _{\alpha })\right) \left\{ \xi _{0\alpha }\,Z\left(
\xi _{n\alpha }\right) -\frac{Z^{\prime }\left( \xi _{n\alpha }\right) }{2}%
\,\left( \frac{T_{\bot \alpha }}{T_{\Vert \alpha }}-\,1\right) \right\} 
\label{11} \\
\varepsilon _{zz} &=&1-\sum_{\alpha }\frac{\omega _{p\alpha }^{2}}{\omega
^{2}}\xi _{0\alpha }\,\sum\limits_{n\,=\,-\,\infty }^{\infty }\Gamma
_{n}(\lambda _{\alpha })\xi _{n\alpha }Z^{\prime }\left( \xi _{n\alpha
}\right) \left\{ 1+\frac{n\Omega _{0\alpha }}{\omega }\left( \frac{T_{\Vert
\alpha }}{T_{\bot \alpha }}-1\right) \right\} \label{12} \\
\varepsilon _{xy} &=&i\sum_{\alpha }\frac{\omega _{p\alpha }^{2}}{\omega ^{2}%
}\sum\limits_{n\,=\,-\,\infty }^{\infty }n\Gamma _{n}^{\prime }(\lambda
_{\alpha })\left\{ \xi _{0\alpha }\,Z\left( \xi _{n\alpha }\right) -\frac{%
Z^{\prime }\left( \xi _{n\alpha }\right) }{2}\,\left( \frac{T_{\bot \alpha }%
}{T_{\Vert \alpha }}-\,1\right) \right\}   \label{13} \\
\varepsilon _{xz} &=&-\sum_{\alpha }\frac{\omega _{p\alpha }^{2}}{\omega ^{2}%
}\frac{v_{t_{\Vert \alpha }}}{v_{t_{\bot \alpha }}}\sum\limits_{n\,=\,-\,%
\infty }^{\infty }\frac{n\Gamma _{n}(\lambda _{\alpha })}{\sqrt{2\lambda
_{\alpha }}}\left\{ \,\xi _{0\alpha }+\,\xi _{n\alpha }\left( \frac{T_{\bot
\alpha }}{T_{\Vert \alpha }}-1\right) \right\} Z^{\prime }\left( \xi
_{n\alpha }\right)  \label{14} \\
\varepsilon _{yz} &=&i\sum_{\alpha }\frac{v_{t_{\Vert \alpha }}}{v_{t_{\bot
\alpha }}}\frac{\omega _{p\alpha }^{2}}{\omega ^{2}}\sum\limits_{n\,=\,-\,%
\infty }^{\infty }\,\sqrt{\frac{\lambda _{\alpha }}{2}}\,\Gamma _{n}^{\prime
}(\lambda _{\alpha })\left\{ \,\xi _{0\alpha }+\,\xi _{n\alpha }\left( \frac{%
T_{\bot \alpha }}{T_{\Vert \alpha }}-1\right) \right\} Z^{\prime }\left( \xi
_{n\alpha }\right)  \label{15}
\end{eqnarray}%
Here, $\epsilon_{ij}$ is the permittivity
dielectric tensor. The parameters in Eqs.~(\ref{10})-(\ref{15}) are
defined as follows:%
\begin{equation}
\xi _{n\alpha }=\frac{\omega \,-n\Omega _{0\alpha }}{k_{\Vert }v_{t\Vert
\alpha }}\,,\,\lambda _{\,\alpha }=\frac{k_{\bot }^{2}v_{t\bot \,\alpha }^{2}%
}{2\Omega _{0\alpha }^{2}}\,,v_{t\Vert \alpha }=\left( 2T_{\Vert \alpha
}/m_{\alpha }\right) ^{1/2}\,\mathrm{and } \quad \,v_{t\bot \alpha }=(2T_{\bot \alpha
}/m_{\alpha })\,^{1/2}  \label{16}
\end{equation}

The results of the integration over $p_{\perp}$ are expressed in terms
of the functions $\Gamma _{n}(\lambda _{\alpha })$ and $\Gamma
_{n}^{\prime }(\lambda _{\alpha })$, which are related to the modified
Bessel function $I_{n}(\lambda _{\,\alpha })$ by the expressions
$\Gamma _{n}(\lambda _{\alpha })=e^{-\lambda _{\,\alpha }}$
$I_{n}(\lambda _{\,\alpha })$ and $\Gamma _{n}^{\prime }(\lambda
_{\alpha })=e^{-\lambda _{\,\alpha }}\left( I_{\,n}^{\prime }(\lambda
  _{\,\alpha })-I_{n}(\lambda _{\,\alpha })\right)$. The integration
over $p_{\parallel}$ introduces the plasma dispersion functions
$Z\left( \xi _{n\alpha }\right) $ and $Z^{\prime }\left( \xi _{n\alpha
  }\right) $ \cite{PDF}. For the isotropic case, the dielectric
permittivity tensor $\epsilon _{ij}$ reduces to the usual textbook
form \cite{chen,brambilla}. 

\subsection{Dispersion Relations for various Modes and Instabilities}
\label{sec:6}
In the limit $\left\vert \xi _{n\alpha }\right\vert \gg 1$,
the general dispersion relation is further simplified, and the
dispersion relations for various modes and instabilities can be
derived. In that limit, the plasma dispersion function assumes the form
\begin{eqnarray}
Z(\xi _{n\alpha }) &=&-\frac{1}{\xi _{n\alpha }}-\frac{1}{2\xi _{n\alpha
}^{3}}-\frac{3}{4\xi _{n\alpha }^{5}}+\cdot \cdot \cdot
=-\sum\limits_{l=0}^{\infty }\frac{\left( 2l+1\right) !!}{\left( 2l+1\right)
\,2^{l}}\left( \frac{k_{\Vert }v_{t_{\Vert \alpha }}}{\omega -n\Omega
_{0\alpha }}\right) ^{2l+1}\mathrm{ for }\left\vert \xi _{n\alpha }\right\vert
\gg 1  \nonumber \label{17}
\end{eqnarray}

For $\left\vert \xi _{n\alpha }\right\vert \gg 1,$ the components of the
tensor in Eq.(\ref{9}) take the form%
\begin{eqnarray}
\epsilon _{xx} &=&1-\sum_{\alpha }\frac{\omega _{p\alpha }^{2}}{\omega ^{2}}%
\sum\limits_{n\,=\,1}^{\infty }\dfrac{n^{2}}{\lambda _{\alpha }}\Gamma
_{n}(\lambda _{\alpha })\sum\limits_{l=0}^{\infty }\frac{\left( 2l+1\right)
!!}{\left( 2l+1\right) }\left\{ \frac{\omega \left( k_{\Vert }v_{t_{\Vert
\alpha }}\right) ^{2l}}{\left( \omega -n\Omega _{0\alpha }\right) ^{2l+1}}+%
\frac{\omega \left( k_{\Vert }v_{t_{\Vert \alpha }}\right) ^{2l}}{\left(
\omega +n\Omega _{0\alpha }\right) ^{2l+1}}\right.  \nonumber \\
&&\ \ \ \ \ \ \ \ \ \ \ \ \ \ \ \ \ \ \ \left. +\frac{\left( 2l+1\right) }{%
2^{l+1}}\left( \frac{T_{\bot \alpha }}{T_{\Vert \alpha }}-1\right) \left(
\left( \frac{k_{\Vert }v_{t_{\Vert \alpha }}}{\omega -n\Omega _{0\alpha }}%
\right) ^{2l+2}+\left( \frac{k_{\Vert }v_{t_{\Vert \alpha }}}{\omega
+n\Omega _{0\alpha }}\right) ^{2l+2}\right) \right\}  \nonumber \\
&&  \label{18} \\
\varepsilon _{yy} &=&1-\left[ \sum_{\alpha }\frac{\omega _{p\alpha }^{2}}{%
\omega ^{2}}\sum\limits_{n\,=\,1}^{\infty }\left( \frac{n^{2}\Gamma
_{n}(\lambda _{\alpha })}{\lambda _{\alpha }}-2\lambda _{\alpha }\,\Gamma
_{n}^{\prime }(\lambda _{\alpha })\right) \sum\limits_{l=0}^{\infty }\frac{%
\left( 2l+1\right) !!}{\left( 2l+1\right) }\left\{ \frac{\omega \left(
k_{\Vert }v_{t_{\Vert \alpha }}\right) ^{2l}}{\left( \omega -n\Omega
_{0\alpha }\right) ^{2l+1}}+\frac{\omega \left( k_{\Vert }v_{t_{\Vert \alpha
}}\right) ^{2l}}{\left( \omega +n\Omega _{0\alpha }\right) ^{2l+1}}\right.
\right.  \nonumber \\
&&\ \ \ \ \ \ \ \ \ \ \ \ \ \ \ \ \ \ \ \left. \left. +\frac{\left(
2l+1\right) }{2^{l+1}}\left( \frac{T_{\bot \alpha }}{T_{\Vert \alpha }}%
-1\right) \left( \left( \frac{k_{\Vert }v_{t_{\Vert \alpha }}}{\omega
-n\Omega _{0\alpha }}\right) ^{2l+2}+\left( \frac{k_{\Vert }v_{t_{\Vert
\alpha }}}{\omega +n\Omega _{0\alpha }}\right) ^{2l+2}\right) \right\} %
\right]  \nonumber \\
&&\ \ \ \ \ \ \ \ \ \ \ \ \ \ \ -2\sum_{\alpha }\frac{\omega _{p\alpha }^{2}%
}{\omega ^{2}}\lambda _{\alpha }\,\Gamma _{0}^{\prime }(\lambda _{\alpha
})\left\{ \xi _{0\alpha }\,Z\left( \xi _{0\alpha }\right) -\frac{Z^{\prime
}\left( \xi _{0\alpha }\right) }{2}\,\left( \frac{T_{\bot \alpha }}{T_{\Vert
\alpha }}-\,1\right) \right\}  \nonumber \\
&&  \label{19} \\
\varepsilon _{zz} &=&1-\sum_{\alpha }\frac{\omega _{p\alpha }^{2}}{\omega
^{2}}\Gamma _{0}(\lambda _{\alpha })\xi _{0\alpha }^{2}\,Z^{\prime }\left(
\xi _{0\alpha }\right) -\sum_{\alpha }\frac{\omega _{p\alpha }^{2}}{\omega }%
\sum\limits_{n\,=\,1}^{\infty }\Gamma _{n}(\lambda _{\alpha
})\sum\limits_{l=0}^{\infty }\frac{\left( 2l+1\right) !!}{2^{l}}\times 
\nonumber \\
\ \ \ &&\ \ \ \ \ \ \ \ \ \ \ \ \ \ \ \times \left[ \frac{\left( k_{\Vert
}v_{t_{\Vert \alpha }}\right) ^{2l}}{\left( \omega -n\Omega _{0\alpha
}\right) ^{2l+1}}+\frac{\left( k_{\Vert }v_{t_{\Vert \alpha }}\right) ^{2l}}{%
\left( \omega +n\Omega _{0\alpha }\right) ^{2l+1}}\right.  \nonumber \\
&&\ \ \ \ \ \ \ \ \ \ \ \ \ \ \ \ \ \ \ \ \ \ \ \ \ \ \ \ \ \ \ \ \ \ \left.
+\frac{n\Omega _{0\alpha }}{\omega }\left( \frac{T_{\Vert \alpha }}{T_{\bot
\alpha }}-1\right) \left( \frac{\left( k_{\Vert }v_{t_{\Vert \alpha
}}\right) ^{2l}}{\left( \omega -n\Omega _{0\alpha }\right) ^{2l+1}}-\frac{%
\left( k_{\Vert }v_{t_{\Vert \alpha }}\right) ^{2l}}{\left( \omega +n\Omega
_{0\alpha }\right) ^{2l+1}}\right) \right]  \nonumber \\
&&  \label{20} \\
\varepsilon _{xy} &=&-\varepsilon _{yx}=-i\sum_{\alpha }\frac{\omega
_{p\alpha }^{2}}{\omega ^{2}}\sum\limits_{n\,=\,1}^{\infty }n\Gamma
_{n}^{\prime }(\lambda _{\alpha })\sum\limits_{l=0}^{\infty }\frac{\left(
2l+1\right) !!}{\left( 2l+1\right) }\left\{ \frac{\omega \left( k_{\Vert
}v_{t_{\Vert \alpha }}\right) ^{2l}}{\left( \omega -n\Omega _{0\alpha
}\right) ^{2l+1}}-\frac{\omega \left( k_{\Vert }v_{t_{\Vert \alpha }}\right)
^{2l}}{\left( \omega +n\Omega _{0\alpha }\right) ^{2l+1}}\right.  \nonumber
\\
&&\ \ \ \ \ \ \ \ \ \ \ \ \ \ \ \ \ \ \ \left. +\frac{\left( 2l+1\right) }{%
2^{l+1}}\left( \frac{T_{\bot \alpha }}{T_{\Vert \alpha }}-1\right) \left(
\left( \frac{k_{\Vert }v_{t_{\Vert \alpha }}}{\omega -n\Omega _{0\alpha }}%
\right) ^{2l+2}-\left( \frac{k_{\Vert }v_{t_{\Vert \alpha }}}{\omega
+n\Omega _{0\alpha }}\right) ^{2l+2}\right) \right\}  \nonumber \\
&&  \label{21}
\end{eqnarray}%
\begin{eqnarray}
\varepsilon _{xz} &=&\varepsilon _{zx}=-\sum_{\alpha }\frac{\omega _{p\alpha
}^{2}}{\omega }\frac{v_{t_{\Vert \alpha }}}{v_{t_{\bot \alpha }}}%
\sum\limits_{n\,=\,1}^{\infty }\frac{n\Gamma _{n}(\lambda _{\alpha })}{\sqrt{%
2\lambda _{\alpha }}}\sum\limits_{l=0}^{\infty }\frac{\left( 2l+1\right) !!}{%
2^{l}}\times  \nonumber \\
&&\ \ \ \ \ \ \ \ \ \ \ \ \ \ \times \left[ \frac{T_{\bot \alpha }}{T_{\Vert
\alpha }}\left\{ \frac{\ \left( k_{\Vert }v_{t_{\Vert \alpha }}\right)
^{2l+1}}{\left( \omega -n\Omega _{0\alpha }\right) ^{2l+2}}-\frac{\ \left(
k_{\Vert }v_{t_{\Vert \alpha }}\right) ^{2l+1}}{\left( \omega +n\Omega
_{0\alpha }\right) ^{2l+2}}\right\} \,\right.  \nonumber \\
&&\ \ \ \ \ \ \ \ \ \ \ \ \ \ \ \ \ \ \ \ \ \ \ \ \left. -\frac{n\Omega
_{0\alpha }}{\omega }\left( \frac{T_{\bot \alpha }}{T_{\Vert \alpha }}%
-1\right) \,\left\{ \frac{\ \left( k_{\Vert }v_{t_{\Vert \alpha }}\right)
^{2l+1}}{\left( \omega -n\Omega _{0\alpha }\right) ^{2l+2}}+\frac{\ \left(
k_{\Vert }v_{t_{\Vert \alpha }}\right) ^{2l+1}}{\left( \omega +n\Omega
_{0\alpha }\right) ^{2l+2}}\right\} \right]  \nonumber \\
&&  \label{22}
\end{eqnarray}%
\begin{eqnarray}
\varepsilon _{yz} &=&-\varepsilon _{zy}=i\sum_{\alpha }\frac{v_{t_{\Vert }}}{%
v_{t\perp }}\frac{\omega _{p\alpha }^{2}}{\omega }\sum\limits_{n\,=\,1}^{%
\infty }\,\sqrt{\frac{\lambda _{\alpha }}{2}}\,\Gamma _{n}^{\prime }(\lambda
_{\alpha })\sum\limits_{l=0}^{\infty }\frac{\left( 2l+1\right) !!\ }{2^{l}}%
\times  \nonumber \\
&&\ \ \ \ \ \ \ \ \ \ \ \times \left[ \frac{T_{\bot \alpha }}{T_{\Vert
\alpha }}\left\{ \frac{\ \left( k_{\Vert }v_{t_{\Vert \alpha }}\right)
^{2l+1}}{\left( \omega -n\Omega _{0\alpha }\right) ^{2l+2}}+\frac{\ \left(
k_{\Vert }v_{t_{\Vert \alpha }}\right) ^{2l+1}}{\left( \omega +n\Omega
_{0\alpha }\right) ^{2l+2}}\right\} \right. \,  \nonumber \\
&&\ \ \ \ \ \ \ \ \ \ \ \ \ \ \ \ \ \ \ \ \ \ \ \left. -\frac{n\Omega
_{0\alpha }}{\omega }\left( \frac{T_{\bot \alpha }}{T_{\Vert \alpha }}%
-1\right) \,\left\{ \frac{\ \left( k_{\Vert }v_{t_{\Vert \alpha }}\right)
^{2l+1}}{\left( \omega -n\Omega _{0\alpha }\right) ^{2l+2}}-\frac{\ \left(
k_{\Vert }v_{t_{\Vert \alpha }}\right) ^{2l+1}}{\left( \omega +n\Omega
_{0\alpha }\right) ^{2l+2}}\right\} \right]  \nonumber \\
&&  \label{23}
\end{eqnarray}%
In Eqs.~(\ref{18})-(\ref{23}), we have separated the $n=0$ terms from
those with $n>0$ and $n<0$. For $n=0,$ we have retained the
plasma dispersion function to be used later. We have also used the
symmetry properties of the modified Bessel function, i~.e., $\Gamma
_{n}(\lambda _{\alpha })=\Gamma _{-n}(\lambda _{\alpha })$ and $\Gamma
_{n}^{\prime }(\lambda _{\alpha })=\Gamma _{-n}^{\prime }(\lambda
_{\alpha })$. Each component of the $\epsilon _{ij}$ tensor
contributes to oblique propagation, since both components of the
wave-number vector, $k_{\bot }$ and $%
k_{\Vert }$ are present via the $\Gamma _{n}(\lambda _{\alpha })$ and
$Z(\xi _{n\alpha })$ functions, respectively, and contain the effects
of thermal anisotropy.

Sections~\ref{sec:1},~\ref{sec:2}~and IIB 3 discuss the parallel, perpendicular and oblique
propagations, respectively, for the limiting case $\left\vert \xi _{n\alpha
}\right\vert \gg 1$.

\subsubsection{Parallel Propagation}
\label{sec:1}
Taking $k_{\bot }=0$ in Eqs.~(\ref{18})-(\ref{23}), we obtain the equations
\begin{eqnarray}
\epsilon _{xx} &=&\varepsilon _{yy}=1-\sum_{\alpha }\frac{\omega _{p\alpha
}^{2}}{2\omega ^{2}}\sum\limits_{l=0}^{\infty }\frac{\left( 2l+1\right) !!}{%
\left( 2l+1\right) }\left\{ \frac{\omega \left( k_{\Vert }v_{t_{\Vert \alpha
}}\right) ^{2l}}{\left( \omega -\Omega _{0\alpha }\right) ^{2l+1}}+\frac{%
\omega \left( k_{\Vert }v_{t_{\Vert \alpha }}\right) ^{2l}}{\left( \omega
+\Omega _{0\alpha }\right) ^{2l+1}}\right.  \nonumber \\
&&\ \ \ \ \ \ \ \ \ \ \ \ \ \ \ \ \ \ \ \left. +\frac{\left( 2l+1\right) }{%
2^{l+1}}\left( \frac{T_{\bot \alpha }}{T_{\Vert \alpha }}-1\right) \left(
\left( \frac{k_{\Vert }v_{t_{\Vert \alpha }}}{\omega -\Omega _{0\alpha }}%
\right) ^{2l+2}+\left( \frac{k_{\Vert }v_{t_{\Vert \alpha }}}{\omega +\Omega
_{0\alpha }}\right) ^{2l+2}\right) \right\}  \nonumber \\
&&  \label{24}\\
\varepsilon _{xy} &=&-\varepsilon _{yx}=-i\sum_{\alpha }\frac{\omega
_{p\alpha }^{2}}{2\omega ^{2}}\sum\limits_{l=0}^{\infty }\frac{\left(
2l+1\right) !!}{\left( 2l+1\right) }\left\{ \frac{\omega \left( k_{\Vert
}v_{t_{\Vert \alpha }}\right) ^{2l}}{\left( \omega -\Omega _{0\alpha
}\right) ^{2l+1}}-\frac{\omega \left( k_{\Vert }v_{t_{\Vert \alpha }}\right)
^{2l}}{\left( \omega +\Omega _{0\alpha }\right) ^{2l+1}}\right.  \nonumber \\
&&\left. +\frac{\left( 2l+1\right) }{2^{l+1}}\left( 
\frac{T_{\bot \alpha }}{T_{\Vert \alpha }}-1\right) \left( \left( \frac{%
k_{\Vert }v_{t_{\Vert \alpha }}}{\omega -\Omega _{0\alpha }}\right)
^{2l+2}-\left( \frac{k_{\Vert }v_{t_{\Vert \alpha }}}{\omega +\Omega
_{0\alpha }}\right) ^{2l+2}\right) \right\}  \label{25} \\
\varepsilon _{zz} &=&1-\sum_{\alpha }\frac{\omega _{p\alpha }^{2}}{\omega
^{2}}\xi _{0\alpha }^{2}\,Z^{\prime }\left( \xi _{o\alpha }\right)
\label{26} \\
\mathrm{and \ \ }\varepsilon _{xz} &=&\varepsilon _{yz}=0  \nonumber
\end{eqnarray}

Here we have expanded the modified Bessel function as follows
\[
\Gamma _{n}(\lambda _{\alpha })=\frac{1}{n!}\left( \frac{\lambda _{\alpha }}{%
2}\right) ^{n},\mathrm{\ }\Gamma _{n}^{\prime }(\lambda _{\alpha })=\frac{n}{%
n!2}\left( \frac{\lambda _{\alpha }}{2}\right) ^{n-1}\left( 1-\frac{\lambda
_{\alpha }}{n}\right) \\ and \quad \Gamma _{0}^{\prime }(\lambda _{\alpha
})=-1
\]
and noted that only the $n=1$ terms survive for $k_{\bot }=0$.\\

Therefore, for parallel propagation, the general dispersion
relation in Eq.(\ref{9}) reduces to the form
\begin{equation}
\left[ \left( \epsilon _{xx}-N_{\parallel }^{2}\right) ^{2}+\epsilon
_{xy}^{2}\right] \epsilon _{zz}=0,  \label{27}
\end{equation}
which decouples the  modes
\begin{equation}
\epsilon _{zz}=0,  \label{28}
\end{equation}
from the electromagnetic modes
\begin{equation}
N_{\parallel }^{2}=\epsilon _{xx}\pm i\epsilon _{xy}.  \label{29}
\end{equation}

The upper ($+$) and the lower ($-$) signs correspond to the R- and L-
waves, respectively. While the dispersion relation for R- and L- waves
retain the thermal anisotropy effect, the electrostatic mode exhibits
no such effect, as expected.\\

\paragraph{R- and L- Waves}$\quad$

$\quad$\\
Using Eqs.(\ref{24})~and (\ref{25}), we may rewrite the dispersion relation
for the R- and L- waves in Eq.(\ref{29}) in the form
\begin{eqnarray}
\frac{c^{2}k_{\Vert }^{2}}{\omega ^{2}} &=&1-\sum_{\alpha }\frac{\omega
_{p\alpha }^{2}}{\omega ^{2}}\sum\limits_{l=0}^{\infty }\frac{\left(
2l+1\right) !!}{\left( 2l+1\right) }\left( \frac{k_{\Vert }v_{t_{\Vert
\alpha }}}{\omega \pm \Omega _{0\alpha }}\right) ^{2l}\left\{ \frac{\omega }{%
\omega \pm \Omega _{0\alpha }}+\frac{\left( 2l+1\right) }{2^{l+1}}\left( 
\frac{T_{\bot \alpha }}{T_{\Vert \alpha }}-1\right) \left( \frac{k_{\Vert
}^{2}v_{t_{\Vert \alpha }}^{2}}{(\omega \pm \Omega _{0\alpha })^{2}}\right)
\right\} \label{30}
\end{eqnarray}

This equation represents the general dispersion relation for R- and L-
waves incorporating the thermal anisotropy as well as the
higher-order thermal effects parallel to the ambient magnetic
field. To retrieve the cold plasma result, we may assume the plasma to
be isotropic and take only the leading term, i.e., $l=0$. Further, with
$\Omega _{0\alpha }=0$ , the structure of both the R- and the L- waves
vanishes altogether and we obtain the dispersion relation for an unmagnetized
plasma. Since the dielectric tensor for more than one species is
additive, we can also solve the dispersion relation for a
multi-species plasma.

The graphical representations of electron R- and L- waves with $l=0$
in Eq.~(30) are shown in Figs.~2 and 3, respectively. The dotted curve
represents the standard isotropic R- and L- wave and the other curves
depict deviation from the isotropic case. We observe that the phase
velocity of the R- and L- waves increases as the magnitude of
anisotropy is increased, but the cutoff point remains unaffected. For
the R-wave, the resonance disappears as anisotropy is switched on.\newline

\paragraph{Whistler Mode }$\quad$\\

$\quad$ If we keep only the leading term (i.e.,
$l=0$) in Eq.(\ref{30}) and assume that $\omega ^{2}\ll c^{2}k_{\Vert
}^{2}$, the dispersion relation for an electron R-wave reduces to
\begin{equation}
\omega =\frac{\Omega _{0e}\,c^{2}k_{\Vert }^{2}}{\left( \omega
_{pe}^{2}+c^{2}k_{\Vert }^{2}\right) }\left\{ 1+\frac{\beta _{\Vert }}{2}%
\left( \frac{T_{\bot e}}{T_{\Vert e}}-1\right) \frac{1}{\left( 1-\dfrac{%
\omega }{\Omega _{0e}}\right) }\right\},  \label{31}
\end{equation}
where 
\begin{equation}
\beta _{\Vert }=\frac{c_{s_{\Vert }}^{2}}{V_{A}^{2}}=\frac{8\pi
n_{o}T_{\Vert e}}{B_{o}^{2}},\mathrm{ with } \quad c_{s_{\Vert }}^{2}=\frac{T_{\Vert
e}}{m_{i}}\,\mathrm{\ ,} \quad V_{A}^{2}=\frac{B_{0}^{2}}{4\pi n_{0}m_{i}}
\label{32}.
\end{equation}

In the low-frequency range $\Omega _{0i}<\omega \,<\Omega _{0e}\,,\ $%
Eq.~(\ref{31}) describes the general whistler mode
\begin{equation}
\omega =\frac{\dfrac{\Omega _{0e}c^{2}k_{\Vert }^{2}}{\omega _{pe}^{2}}%
\left\{ 1+\left( \dfrac{T_{\bot e}}{T_{\Vert e}}-1\right) \dfrac{\beta
_{\Vert }}{2}\right\} }{\left\{ \left( \dfrac{c^{2}k_{\Vert }^{2}}{\omega
_{pe}^{2}}+1\right) -\left( \dfrac{T_{\bot e}}{T_{\Vert e}}-1\right) \dfrac{%
k_{\Vert }^{2}v_{t_{\Vert e}}^{2}}{2\Omega _{0e}^{2}}\right\} }.  \label{33}
\end{equation}
Here we have retained the first-order term in $\dfrac{\omega }{\Omega _{0e}}$.\\

For longer wavelengths, i.e., $\omega _{pe}^{2}\gg c^{2}k_{\Vert }^{2}$,
this dispersion relation reduces to the form
\begin{equation}
\omega =\frac{\dfrac{\Omega _{0e}c^{2}k_{\Vert }^{2}}{\omega _{pe}^{2}}%
\left\{ 1+\left( \dfrac{T_{\bot e}}{T_{\Vert e}}-1\right) \dfrac{\beta
_{\Vert }}{2}\right\} }{\left\{ 1-\left( \dfrac{T_{\bot e}}{T_{\Vert e}}%
-1\right) \dfrac{k_{\Vert }^{2}v_{t_{\Vert e}}^{2}}{2\Omega _{0e}^{2}}%
\right\} }.  \label{34}
\end{equation}

This is the dispersion relation for a whistler mode including the
electronic thermal anisotropy effect. If we neglect the second order
term in the denominator on the right-hand side, which is due to the first order
term in $\dfrac{\omega }{\Omega _{0e}}$, we recover the results of
Lazar et al.,\ \cite{lazar}. Figure~4 shows how the phase speed of the
whistler wave is enhanced as the thermal anisotropy increases.\newline

\paragraph{Non-Resonant Whistler Instability ($\left\vert \xi_{n\alpha }\right\vert \gg 1$)} $\quad$

$\quad$\\
By considering only the leading term (i.e., $l=0$) in the R-wave and
assuming that $\omega ^{2}\ll c^{2}k_{\Vert }^{2}$ in
Eq.~(\ref{30})$,$ we obtain the following expressions for the
real and imaginary parts of $\omega$:
\begin{equation}
\Re\omega =\Omega _{0e}\left( 1-\frac{\mathbf{\,}\omega _{pe}^{2}}{%
2\left( c^{2}k_{\Vert }^{2}\mathbf{+}\omega _{pe}^{2}\right) }\right)
\label{35}
\end{equation}
and
\begin{equation}
\Im\omega =\omega _{pe}\frac{\sqrt{2\left( \frac{T_{\bot e}}{T_{\Vert
e}}-1\right) k_{\Vert }^{2}v_{t_{\Vert e}}^{2}\left( c^{2}k_{\Vert }^{2}%
\mathbf{+}\omega _{pe}^{2}\right) -\Omega _{0e}^{2}\omega _{pe}^{2}}}{%
2\left( c^{2}k_{\Vert }^{2}\mathbf{+}\omega _{pe}^{2}\right) }  \label{36}
\end{equation}%
in agreement with Lee et al.,\ \cite{lee} and Lazar et al., \ \cite{lazar}.

The instability occurs for wave numbers satisfying the condition
\begin{equation}
k_{\Vert }^{2}>k_{m}^{2}=\frac{\omega _{pe}^{2}}{2\,c^{2}}\left[ \left\{ 1+%
\frac{2}{\left( \frac{T_{\bot e}}{T_{\Vert e}}-1\right) }\left( \frac{\Omega
_{0e}^{2}}{\omega _{pe}^{2}}\right) \left( \frac{c^{2}}{v_{t_{\Vert e}}^{2}}%
\right) \right\} ^{\frac{1}{2}}-1\right].  \label{37}
\end{equation}

From Fig.~5, we see that the growth rate increases as the thermal
anisotropy increases, while the instability threshold is lowered. For any specific anisotropy value, the growth rate initially rises for small $k_{\Vert }$
and then saturates at large $k_{\Vert }$.

From Eq.~(\ref{36}), we see that the ambient magnetic field
suppresses the growth rate. In the limit of large thermal
anisotropy, i.e., for $\frac{T_{\bot }}{T_{\Vert }}\gg 1$, the dispersion
relation reduces to
\begin{equation}
\Im\omega =\omega _{pe}\frac{\sqrt{2k_{\Vert }^{2}v_{t_{\bot
e}}^{2}\left( c^{2}k_{\Vert }^{2}\mathbf{+}\omega _{pe}^{2}\right) -\Omega
_{0e}^{2}\omega _{pe}^{2}}}{2\left( c^{2}k_{\Vert }^{2}\mathbf{+}\omega
_{pe}^{2}\right) }.  \label{38}
\end{equation}

If the magnetic field is neglected, Eq.~(\ref{38}) reduces to the
simpler expression
\[
\Im\omega =\frac{k_{\Vert }v_{t_{\bot e}}}{\sqrt{2}}\frac{\omega _{pe}%
}{\sqrt{\left( c^{2}k_{\Vert }^{2}\mathbf{+}\omega _{pe}^{2}\right) }} 
\]
which is the well-known result due to Weibel \cite{weibel}.\newline

\paragraph{Alfv\'{e}n Wave} $\quad$

$\quad$\\
We now consider the dispersion relation~(\ref{30}) for a two-component plasma
consisting of electrons and singly charged ions. In the low-frequency
regime $\omega <\Omega _{i}$, we keep only the leading term, i.e., $l=0,$
to rewrite Eq.~(\ref{30}) in the form
\begin{eqnarray}
\frac{\omega ^{2}}{k_{\Vert }^{2}V_{A}^{2}} &=&\left( \frac{c^{2}}{%
V_{A}^{2}+c^{2}}\right) \left[ 1+\frac{\beta _{\Vert }}{2}\left\{ \left( 
\frac{T_{\bot e}}{T_{\Vert }}-1\right) \left( 1\pm \dfrac{2\omega }{\Omega
_{0e}}\right) +\left( \frac{T_{\bot i}}{T_{\Vert }}-1\right) \left( 1\mp 
\dfrac{2\omega }{\Omega _{0i}}\right) \right\} \right]  \nonumber \\
&&  \label{39}
\end{eqnarray}
Here we have assumed that the two species are at the same parallel temperature
and taken advantage of the small mass ratio, $m_{e}/m_{i}\ll 1$.

In the isotropic case, Eq.~(\ref{39}) reduces to the standard pure
Alfv\'{e}n mode. For frequencies satisfying $\omega\ll \Omega _{0i}$,
the quadratic structure in $\omega $ vanishes, and both the R- and the L-
waves follow the same dispersion relation,
\begin{equation}
\omega =k_{\Vert }V_{A}\left[ 1+\frac{\beta _{\Vert }}{2}\left\{ \left( 
\frac{T_{\bot e}}{T_{\Vert }}-1\right) +\left( \frac{T_{\bot i}}{T_{\Vert }}%
-1\right) \right\} \right] ^{\dfrac{1}{2}},  \label{40}
\end{equation}
a result derived by Schlickeiser and Skoda \cite{p. Alfven}.

In Eqs.~(\ref{39})~and (\ref{40}), the Alfv\'{e}n wave frequency is
modified by acoustic effects due to the thermal anisotropy of
the electrons and ions. The Alfv\'{e}n phase velocity can be enhanced or reduced,
depending on the intensity and signatures of the anisotropy.

From Eq.(\ref{40}), it is evident that the constraint 
\[
\frac{2}{\beta _{\Vert }}\ll \left( 1-\frac{T_{\bot e}}{T_{\Vert }}\right)
+\left( 1-\frac{T_{\bot i}}{T_{\Vert }}\right) 
\]
makes the Alfv\'{e}n mode unstable, for then
\begin{equation}
\omega =i\,k_{\Vert }V_{A}\left[ \frac{\beta _{\Vert }}{2}\left\{ \left( 1-%
\frac{T_{\bot e}}{T_{\Vert }}\right) +\left( 1-\frac{T_{\bot i}}{T_{\Vert }}%
\right) \right\} -1\right] ^{\dfrac{1}{2}},  \label{41}
\end{equation}
which describes the so-called fire-hose instability. The Alfv\'en-wave
frequency is real for $T_{\bot e,i}>T_{\Vert e,i}$, under which
condition fire-hose instabilities cannot arise.\newline

\paragraph{Electrostatic Waves} $\quad$

$\quad$\\
From Eq.~(\ref{28}), i.e., $\epsilon _{zz}=0$, we obtain the dispersion
relation for Langmuir waves
\begin{equation}
\omega ^{2}=\omega _{pe}^{2}+\frac{3}{2}k_{\Vert }^{2}v_{t_{\Vert e}}^{2},
\label{42}
\end{equation}
where we have used the condition $\left\vert \xi _{0e}\right\vert \gg
1$ for expanding the plasma dispersion function $Z(\xi _{oe})$.

For $v_{t\Vert i}\ll \omega \,/k_{\Vert }\ll v_{t\Vert e\,}$ and $k_{\Vert
}^{2}\lambda _{De}^{2}\ll 1,$ the equation $\epsilon _{zz}=0$ yields the
ion-acoustic mode
\begin{equation}
\omega ^{2}=k_{\Vert }^{2}c_{s_{\Vert }}^{2}  \label{43}
\end{equation}
where we have used that
\[
Z^{\prime }\left( \xi _{oe}\right) =-2 \quad \mathrm{ for } \quad \left\vert \xi
_{oe}\right\vert \ll 1\mathrm{, } \quad Z^{\prime }\left( \xi _{oi}\right) =\frac{1}{%
\xi _{oi}^{2}} \quad \mathrm{ \thinspace for } \quad \left\vert \xi _{oi}\right\vert \gg 1%
\quad \mathrm{ and } \quad c_{s_{\Vert }}^{2}=\frac{T_{\Vert e}}{m_{i}} 
\]

The electrostatic modes, which are only affected by the temperature in
the direction of propagation, are hence insensitive to the thermal
anisotropy.

\subsubsection{Perpendicular Propagation}
\label{sec:2}
Letting $k_{\Vert }=0$ and noting that only the leading terms in
the $l$-summation (i.e., $l=0)$ survive in Eqs.~(\ref{18})-(\ref{23}), we
obtain the expressions
\begin{eqnarray}
\varepsilon _{xx} &=&1-\sum_{\alpha }\frac{\omega _{p\alpha }^{2}}{\omega ^{2}}%
\sum\limits_{n\,=\,1}^{\infty }\dfrac{n^{2}}{\lambda _{\alpha }}\Gamma
_{n}(\lambda _{\alpha })\left( \frac{2\omega ^{2}}{\omega ^{2}-n^{2}\Omega
_{0\alpha }^{2}}\right)  \label{44} \\
\varepsilon _{yy} &=&1-\sum_{\alpha }\frac{\omega _{p\alpha }^{2}}{\omega
^{2}}\sum\limits_{n\,=\,1}^{\infty }\left( \frac{n^{2}\Gamma _{n}(\lambda
_{\alpha })}{\lambda _{\alpha }}-2\lambda _{\alpha }\,\Gamma _{n}^{\prime
}(\lambda _{\alpha })\right) \left( \frac{2\omega ^{2}}{\omega
^{2}-n^{2}\Omega _{0\alpha }^{2}}\right) +2\sum_{\alpha }\frac{\omega
_{p\alpha }^{2}}{\omega ^{2}}\lambda _{\alpha }\,\Gamma _{0}^{\prime
}(\lambda _{\alpha })  \label{45} \\
\varepsilon _{zz} &=&1-\sum_{\alpha }\frac{\omega _{p\alpha }^{2}}{\omega
^{2}}\Gamma _{0}(\lambda _{\alpha })-2\sum_{\alpha }\frac{\omega _{p\alpha
}^{2}}{\omega ^{2}}\sum\limits_{n\,=\,1}^{\infty }\Gamma _{n}(\lambda
_{\alpha })\left\{ 1+\frac{T_{\Vert \alpha }}{T_{\bot \alpha }}\frac{%
n^{2}\Omega _{0\alpha }^{2}}{\omega ^{2}-n^{2}\Omega _{0\alpha }^{2}}\right\}
\label{46} \\
\varepsilon _{xy} &=&-\varepsilon _{yx}=-i\sum_{\alpha }\frac{\omega
_{p\alpha }^{2}}{\omega ^{2}}\sum\limits_{n\,=\,1}^{\infty }n\Gamma
_{n}^{\prime }(\lambda _{\alpha })\left( \frac{2n\omega \Omega _{0\alpha }}{%
\omega ^{2}-n^{2}\Omega _{0\alpha }^{2}}\right)  \label{47} \\
\varepsilon _{xz} &=&\varepsilon _{yz}=0  \nonumber
\end{eqnarray}
In the expressions for $\epsilon _{yy}$ and $\epsilon _{zz}$,
Eqs.~(\ref{19})~and (\ref{20}), respectively, we have expanded the plasma-dispersion
functions $Z(\xi _{0\alpha })$ and $Z^{\prime }(\xi _{0\alpha })$ for
$\left\vert \xi _{0\alpha }\right\vert \gg 1$.

Thus the general dispersion relation in Eq.~(\ref{9}) reduces to
\begin{equation}
\left[ \epsilon _{xx}\left( \epsilon _{yy}-N_{\bot }^{2}\right) +\epsilon
_{xy}^{2}\right] \left( \epsilon _{zz}-N_{\perp }^{2}\right) =0  \label{48}
\end{equation}

The above equation shows two decoupled modes: the Extraordinary (X-)
mode
\begin{equation}
N_{\bot }^{2}=\epsilon _{yy}\left( 1+\frac{\epsilon _{xy}^{2}}{\epsilon
_{xx}\epsilon _{yy}}\right)  \label{49}
\end{equation}
and the Ordinary (O-) mode
\begin{equation}
N_{\perp }^{2}=\epsilon _{zz}  \label{50}
\end{equation}

If we assume the non-diagonal components $\epsilon _{xy}$ to be much
smaller than the diagonal components $\epsilon_{xx}$ and
$\epsilon_{yy},$ the X-mode dispersion relation~(\ref{49}) yields two
decoupled modes: the purely transversal X-mode
\begin{equation}
N_{\bot }^{2}=\epsilon _{yy},  \label{51}
\end{equation}
and the Bernstein mode
\begin{equation}
\epsilon _{xx}=0.  \label{52}
\end{equation}

Here, like in parallel propagation, we observe that the thermal
anisotropy affects the dispersion relation of the O-mode
[Eq.(\ref{50})], but does not affect the X-mode [Eq.(\ref{49})]
or the Bernstein mode [Eq.(\ref{52})].

We will next discuss these three modes in more detail.\newline

\paragraph{X- Mode} $\quad$

$\quad$\\
With the help of Eqs.~(\ref{44}), (\ref{45})~ and (\ref{47}) we
rewrite the X-mode dispersion relation, Eq.~(\ref{49}) in the form
\begin{eqnarray}
\frac{c^{2}k_{\bot }^{2}}{\omega ^{2}} &=&\left[ 1-\sum_{\alpha }\frac{%
\omega _{p\alpha }^{2}}{\omega ^{2}}\sum\limits_{n\,=\,1}^{\infty }\left( 
\frac{n^{2}\Gamma _{n}(\lambda _{\alpha })}{\lambda _{\alpha }}-2\lambda
_{\alpha }\,\Gamma _{n}^{\prime }(\lambda _{\alpha })\right) \left( \frac{%
2\omega ^{2}}{\omega ^{2}-n^{2}\Omega _{0\alpha }^{2}}\right) \right] 
\nonumber \\
&&-\left( \frac{\left[ \sum_{\alpha }\dfrac{\omega _{p\alpha }^{2}}{\omega
^{2}}\sum\limits_{n\,=\,1}^{\infty }n\Gamma _{n}^{\prime }(\lambda _{\alpha
})\left( \dfrac{2n\omega \Omega _{0\alpha }}{\omega ^{2}-n^{2}\Omega
_{0\alpha }^{2}}\right) \right] ^{2}}{\left[ 1-\sum_{\alpha }\dfrac{\omega
_{p\alpha }^{2}}{\omega ^{2}}\sum\limits_{n\,=\,1}^{\infty }\dfrac{n^{2}}{%
\lambda _{\alpha }}\Gamma _{n}(\lambda _{\alpha })\left( \dfrac{2\omega ^{2}%
}{\omega ^{2}-n^{2}\Omega _{0\alpha }^{2}}\right) \right] }\right) \label{53},
\end{eqnarray}
which is the general form of the X-mode. 

Figure~6 depicts the behavior of X-mode for an electron plasma.
Panels (a), (c), and (e) show the behaviors for $n=1$, and the
harmonics $n=2$, and 3, respectively, with $\lambda_e=0.05$, while
panels (b), (d), and (f) correspond to $n=1$, 2, and 3, respectively,
with $\lambda_e=8.0$.  The number of cutoffs and resonances increases
with $n$. The distinction between 
the resonance and the cutoff points is more prominent in the right
panels, i.e., for the large argument. The cutoff and
resonance points shift towards lower frequencies and hence expands the
propagation domain as $\lambda_e$ grows from 0.05 to 8.0.

The pure transverse X-mode (i.e.,$c^{2}k_{\bot }^{2}/\omega ^{2}
=\epsilon _{yy}$), given by Eq.(\ref{51}), may be written as
\begin{equation}
\frac{c^{2}k_{\bot }^{2}}{\omega ^{2}}=1-\sum_{\alpha }\frac{\omega
_{p\alpha }^{2}}{\omega ^{2}}\sum\limits_{n\,=\,1}^{\infty }\left( \frac{%
n^{2}\Gamma _{n}(\lambda _{\alpha })}{\lambda _{\alpha }}-2\lambda _{\alpha
}\,\Gamma _{n}^{\prime }(\lambda _{\alpha })\right) \left( \frac{2\omega ^{2}%
}{\omega ^{2}-n^{2}\Omega _{0\alpha }^{2}}\right) +2\sum_{\alpha }\frac{%
\omega _{p\alpha }^{2}}{\omega ^{2}}\lambda _{\alpha }\,\Gamma _{0}^{\prime
}(\lambda _{\alpha })  \label{54}
\end{equation}

The two X-mode dispersion relations, Eqs.~(\ref{53})~and (\ref{54}),
contain the standard modified Bessel function. We can solve these
relations numerically to describe the X-mode in detail. Alternatively,
we can seek an approximate analytical solution results by expanding
the Bessel function for small argument, which converts
Eqs.~(\ref{53}),~and (\ref{54}) to the form
\begin{eqnarray}
\dfrac{c^{2}k_{\bot }^{2}}{\omega ^{2}} &=&\left\{ 1-\sum_{\alpha }\frac{%
\omega _{p\alpha }^{2}}{\omega ^{2}}\sum\limits_{n\,=\,1}^{\infty }\frac{%
n^{2}}{n!}\left( \frac{k_{\bot }v_{t\bot }}{2\Omega _{0\alpha }}\right)
^{2n-2}\left( \frac{\omega ^{2}}{\omega ^{2}-n^{2}\Omega _{0\alpha }^{2}}%
\right) \,\right\}  \nonumber \\
&&  \nonumber \\
&&-\left[ \dfrac{\left\{ \sum_{\alpha }\dfrac{\omega _{p\alpha }^{2}}{\omega
^{2}}\sum\limits_{n\,=\,1}^{\infty }\frac{n^{2}}{n!}\left( \dfrac{k_{\bot
}v_{t\bot }}{2\Omega _{0\alpha }}\right) ^{2n-2}\left( \dfrac{n\omega \Omega
_{0\alpha }}{\omega ^{2}-n^{2}\Omega _{0\alpha }^{2}}\right) \right\} ^{2}}{%
\left\{ 1-\sum_{\alpha }\dfrac{\omega _{p\alpha }^{2}}{\omega ^{2}}%
\sum\limits_{n\,=\,1}^{\infty }\frac{n^{2}}{n!}\left( \dfrac{k_{\bot
}v_{t\bot }}{2\Omega _{0\alpha }}\right) ^{2n-2}\left( \dfrac{\omega ^{2}}{%
\omega ^{2}-n^{2}\Omega _{0\alpha }^{2}}\right) \right\} }\right],  \label{55}
\end{eqnarray}
and
\begin{equation}
\frac{c^{2}k_{\bot }^{2}}{\omega ^{2}}=1-\sum_{\alpha }\frac{\omega
_{p\alpha }^{2}}{\omega ^{2}}\sum\limits_{n\,=\,1}^{\infty }\frac{n^{2}}{n!}%
\left( \frac{k_{\bot }v_{t\bot }}{2\Omega _{0\alpha }}\right) ^{2n-2}\left( 
\frac{\omega ^{2}}{\omega ^{2}-n^{2}\Omega _{0\alpha }^{2}}\right)
\label{56},
\end{equation}
respectively.

Equation~(\ref{56}) is the expression derived by Zaheer and Murtaza
for pure transverse X-modes \cite{sadia4}. Equation~(\ref{55})
with the sums over $n$ on the right-hand side truncated at $n=1$ reduces
to the textbook expression for the dispersion relation for the
general X-mode in electron plasmas (see, e.~g., Chen \cite{chen})
:
\begin{equation}
\frac{c^{2}k_{\bot }^{2}}{\omega ^{2}}=1-\frac{\omega _{pe}^{2}}{\omega ^{2}}%
\left( \frac{\omega ^{2}-\omega _{pe}^{2}}{\omega ^{2}-(\omega
_{pe}^{2}+\Omega _{0e}^{2})}\right)  \label{57}
\end{equation}\newline

\paragraph{Bernstein Wave} $\quad$

$\quad$\\
In view of the expression for $\epsilon _{xx}$, Eq.~(\ref{44}), the dispersion
relation for the Bernstein wave $\epsilon _{xx}=0$ in Eq.~(\ref{52}) takes the
form
\begin{equation}
1=2\sum_{\alpha }\frac{\omega _{p\alpha }^{2}}{\omega ^{2}}\left( \frac{\exp
[-\lambda _{\,\alpha }]}{\lambda _{\alpha }}\right)
\sum\limits_{n\,=\,1}^{\infty }n^{2}I_{n}(\lambda _{\,\alpha })\left( \frac{%
\omega ^{2}}{\omega ^{2}-n^{2}\Omega _{0\alpha }^{2}}\right),  \label{58}
\end{equation}
Figure~7 depicts solutions of this equation. The
dotted line represents the solution for $\lambda _{e}=1$ and the other
three curves represent $\lambda_{e}= 0.5$, 1.5, and 2.0. The resonance points 
are fixed, but the cutoff points shift from right to left and reduce
the propagation domain as $\lambda _{e}$ grows. At higher harmonics
(not shown), the cutoff points become independent of $\lambda_{e}$.\newline

\paragraph{O- Mode} $\quad$

$\quad$\\
Given the expression for $\epsilon _{zz}$, Eq.~(\ref{46}), the dispersion
relation for the O-mode in Eq.(\ref{50}) is given by the equality
\begin{eqnarray}
\frac{c^{2}k_{\bot }^{2}}{\omega ^{2}} &=&1-\sum_{\alpha }\frac{\omega
_{p\alpha }^{2}}{\omega ^{2}}\Gamma _{0}(\lambda _{\alpha })-2\sum_{\alpha }%
\frac{\omega _{p\alpha }^{2}}{\omega ^{2}}\sum\limits_{n\,=\,1}^{\infty
}\Gamma _{n}(\lambda _{\alpha })\left\{ 1+\frac{T_{\Vert \alpha }}{T_{\bot
\alpha }}\frac{n^{2}\Omega _{0\alpha }^{2}}{\omega ^{2}-n^{2}\Omega
_{0\alpha }^{2}}\right\} \label{59}
\end{eqnarray}

Again we expand the Bessel function for small argument and rewrite
this equation in the form
\begin{eqnarray}
\frac{c^{2}k_{\bot }^{2}}{\omega ^{2}} &=&1-\sum_{\alpha }\frac{\omega
_{p\alpha }^{2}}{\omega ^{2}}-2\sum_{\alpha }\frac{\omega _{p\alpha }^{2}}{%
\omega ^{2}}\sum\limits_{n\,=\,1}^{\infty }\frac{1}{n!}\left( \frac{k_{\bot
}v_{t\bot }}{2\Omega _{0\alpha }}\right) ^{2n}\left\{ \frac{\omega ^{2}}{%
\omega ^{2}-n^{2}\Omega _{0\alpha }^{2}}-\left( 1-\frac{T_{\Vert \alpha }}{%
T_{\bot \alpha }}\right) \frac{n^{2}\Omega ^{2}}{\omega ^{2}-n^{2}\Omega
_{0\alpha }^{2}}\right\} \label{60}
\end{eqnarray}
This is a general dispersion relation for the O-mode with higher order
thermal effects, including the consequences of thermal anisotropy.

\subsubsection{Oblique Propagation}
\label{sec:7}
In the low-frequency, long parallel-wavelength regime, the non-diagonal components
of the tensor $\varepsilon _{ij}$ are negligibly small. Therefore the
dispersion relation in Eq.~(9), for the kinetic Alfv\'{e}n waves (KAWs) can be
written as 
\begin{equation}
\left\vert 
\begin{array}{cc}
\varepsilon _{xx}-N_{\Vert }^{2} & N_{\Vert }N_{\bot } \\ 
N_{\Vert }N_{\bot } & \varepsilon _{zz}-N_{\bot }^{2}%
\end{array}%
\right\vert =0.  \label{61}
\end{equation}

In this regime, the fast mode is described by the expression
\begin{equation}
\varepsilon _{yy}-N^{2}=0,  \label{62}
\end{equation}
and is hence decoupled from the KAWs.

The electric field vector $\mathbf{E}$ and the wavenumber vector $\mathbf{k}$
are coplanar for the KAWs \cite{29}-\cite{33}.

Reference~\cite{Bashir} presents a detailed derivation of the
expressions describing the oblique propagation of the KAWs in the
kinetic and inertial limits. Here we present a simpler derivation.  To
highlight the effects of thermal anisotropy and to show the wave
frequency $\omega$ as a function of the perpendicular and parallel
wave vectors $k_{\Vert}$ and $k_{\bot }$, we present a 3-D graphical
representation of the simplified results. To the best of our
knowledge, our results for the general fast mode and the fast mode
instability are new.

General expressions for the components of the tensor $\varepsilon
_{ij}$ were presented in Section~\ref{sec:4}. We substitute
Eq.~\eqref{10} for $\varepsilon_{xx}$ and Eq.~\eqref{12} for
$\varepsilon_{zz}$ and consider
the low-frequency limit $\omega \ll \Omega _{i}$ , $k_{\Vert
}^{2}\lambda _{De}^{2}\ll 1$, $V_{A}^{2}$ $\ll c^{2}$ to reduce
Eq.~\eqref{61} to the equality
\begin{eqnarray}
&&\left( \frac{\omega ^{2}\,}{k_{_{\Vert }}^{2}\,V_{A}^{2}\,\,\,\,}-\dfrac{%
(1+\psi _{1})\lambda _{i}}{1-\Gamma _{0}(\lambda _{i})}\right) \left( -\frac{%
\Gamma _{0}(\lambda _{e})\,}{2k_{\Vert }^{2}\lambda _{De}^{2}}Z^{\prime
}\left( \xi _{0e}\right) -\frac{\Gamma _{0}(\lambda _{i})\,}{2k_{\Vert
}^{2}\lambda _{Di}^{2}}\,Z^{\prime }\left( \xi _{0i}\right) +\frac{\omega
_{pi}^{2}}{\omega ^{2}}\psi _{2}\right)  \nonumber \\
&&=\left( \frac{\omega ^{2}\,}{k_{_{\Vert }}^{2}\,V_{A}^{2}\,\,\,\,}-\left( 
\dfrac{\lambda _{i}}{1-\Gamma _{0}(\lambda _{i})}\right) \psi _{1}\right) 
\frac{c^{2}k_{\bot }^{2}\,\,\,}{\,\omega ^{2}\,\,}  \label{63}
\end{eqnarray}
with the shorthand
\[
\lambda _{D\alpha }^{2}=\frac{v_{t\Vert \alpha }^{2}}{2\omega _{p\alpha }^{2}%
}.
\]

Here, the anisotropy terms $\psi _{1}$ and $\psi _{2}$ are
\begin{equation}
\psi _{1}=\frac{c_{s_{\Vert }}^{2}}{V_{A}^{2}\,}\left\{ \left( \dfrac{%
1-\Gamma _{0}(\lambda _{i})}{\lambda _{i}}\,\right) \left( \frac{T_{\bot i}}{%
T_{\Vert i}}-\,1\right) \left( \frac{T_{\Vert i}}{T_{\Vert e}}\right)
+\left( \dfrac{1-\Gamma _{0}(\lambda _{e})}{\lambda _{e}}\right) \,\left( 
\frac{T_{\bot e}}{T_{\Vert e}}-\,1\right) \right\},  \label{64}
\end{equation}
and
\begin{equation}
\psi _{2}=\left\{ \frac{m_{i}}{m_{e}}\,\left( 1-\Gamma _{0}(\lambda
_{e})\right) \left( \frac{T_{\Vert e}}{T_{\bot e}}-1\right) +\,\left(
1-\Gamma _{0}(\lambda _{i})\right) \left( \frac{T_{\Vert i}}{T_{\bot i}}%
-1\right) \right\}  \label{65}
\end{equation}

Although in Eq.~(\ref{63}),  we have included contributions from both species for terms containing anisotropic effect but we
have neglected the electronic contribution from the isotropic part of
the $\epsilon _{xx}$ component by using the small mass ratio,
$m_{e}/m_{i}\ll 1$.

Next, we shall derive the dispersion relation for the kinetic
Alfv\'{e}n wave in the kinetic (i.e., $v_{t\Vert i}\ll \omega
\,/k_{\Vert }\ll v_{t\Vert e\,}$ and $(m_{e}/m_{i})\ll \beta
  =(c_{s_{\Vert }}^{2}/V_{A}^{2})\ll 1)$  and in the inertial
  ($v_{t\Vert e,i}\ll \omega \,/k_{\Vert }$ , $\beta \ll m_{e}/m_{i}$)
  limits. The kinetic (inertial) limit is defined with reference to the
  dominant thermal (inertial) electronic effect.\newline

\paragraph{KAWs in the Kinetic Limit} $\quad$

$\quad$\\
To discuss the kinetic limit of KAWs, we assume the parallel phase velocity
of the wave to be less than the parallel thermal velocity
of the electrons, yet greater than the parallel thermal velocity of
the ions, i.e., $v_{t\Vert i}\ll \omega
\,/k_{\Vert }\ll v_{t\Vert e\,}$, and the plasma to have low
$\beta$,  i.e., $(m_{e}/m_{i})\ll \beta =(c_{s_{\Vert }}^{2})/V_{A}^{2})\ll 1$.
Under these conditions, the dispersion relation in Eq.~(\ref{63}) reduces to
\begin{equation}
\omega ^{2}=k_{_{\Vert }}^{2}\,V_{A}^{2}\,\left\{ \,1+\frac{3}{4}k_{\bot
}^{2}\rho _{i}^{2}+k_{\bot }^{2}\rho _{s}^{2}\,\,+\psi _{1}^{\prime }\,\frac{%
c_{s_{\Vert }}^{2}}{V_{A}^{2}\,}\right\},  \label{66}
\end{equation}
where
\begin{equation}
\psi _{1}^{\prime }=\left\{ \left( 1-\frac{3}{4}k_{\bot }^{2}\rho
_{e}^{2}\right) \left( \dfrac{T_{\bot e}}{T_{\parallel e\,}}-1\right) +%
\dfrac{T_{\parallel i\,}}{T_{\parallel e\,}}\left( 1-\frac{3}{4}k_{\bot
}^{2}\rho _{i}^{2}\right) \left( \dfrac{T_{\bot i}}{T_{\parallel i\,}}%
-1\right) \right\},  \label{67}
\end{equation}
and 
\[
\rho _{s}^{2}=\frac{c_{s_{\Vert }}^{2}}{\Omega _{0i}^{2}}
\]

In deriving Eq.(\ref{66}) we have assumed the gyro-radii to be small,
so that
\begin{equation}
\Gamma _{0}(\lambda _{e,i})\approx 1-\lambda _{e,i}+\frac{3}{4}\lambda
_{e,i}^{2} \quad \mathrm{ and } \quad \dfrac{\lambda _{e,i}}{1-\Gamma _{0}(\lambda _{e,i})}=%
\dfrac{1}{1-\frac{3}{4}k_{\bot }^{2}\rho _{e,i}^{2}}\simeq 1+\frac{3}{4}%
k_{\bot }^{2}\rho _{e,i}^{2}.  \label{68}
\end{equation}

Equation~(\ref{63}) describes the general kinetic Alfv\'{e}n wave,
in which $\psi_{1}^{\prime }$ measures the deviation from the
isotropic limit. Illustrative plots are presented in Fig.~8.\newline

\paragraph{\ KAWs in the Inertial Limit} $\quad$

$\quad$\\
To obtain the inertial limit of KAWs, we assume that the parallel
phase velocity of the wave is greater than the parallel thermal
velocities of both the electrons and the ions, i.e., $v_{t\Vert
  e,i}\ll \omega \,/k_{\Vert }$ , $\beta \ll (m_{e}/m_{i})$, and
that the gyroradii are small.  Under these constraints, the
dispersion relation in Eq.~(\ref{63}) yields the following expression
for the modified kinetic Alfv\'{e}n wave in the inertial regime:
\begin{equation}
\omega ^{2}=k_{\parallel }^{2}V_{A}^{2}\left[ \frac{1}{1+\,\dfrac{%
c^{2}k_{\bot }^{2}}{\omega _{pe}^{2}}}+\beta _{\Vert }\left\{ \dfrac{%
T_{\parallel i\,}}{T_{\parallel e\,}}\left( \dfrac{T_{\bot e}}{T_{\parallel
e\,}}-1\right) +\left( \dfrac{T_{\bot i}}{T_{\parallel i\,}}-1\right)
\right\} \right].  \label{69}
\end{equation}

The inertial Alfv\'{e}n wave is modified by the acoustic effect, in
turn caused by the thermal anisotropy. Since $\beta\ll m_{e}/m_{i}$
only extraordinarily large thermal anisotropies will show appreciable
effects.\newline

\paragraph{Fast Mode} $\quad$

$\quad$\\
We take advantage of the low-frequency limit $\omega \ll \Omega
_{0\alpha }$ and of the Bessel function identity $2\sum\limits_{n\,=
  \,1}^{\infty }\Gamma _{n}(\lambda _{\alpha })=1-\Gamma _{0}(\lambda
_{\alpha })$ and keep only the leading term ($l=0$) in the sum over
$l$ on the right-hand side of Eq.~(\ref{21}). The fast mode dispersion
relation in Eq.~(\ref{62}) ( i.e., $\varepsilon _{yy}-N^{2}=0$) then reduces
to
\begin{eqnarray}
\frac{c^{2}k^{2}}{\omega ^{2}} &=&1+\sum_{\alpha }\frac{\omega _{p\alpha
}^{2}}{\Omega _{\alpha }^{2}}\left( \frac{1-\Gamma _{0}(\lambda _{\alpha })}{%
\lambda _{\alpha }}-4\sum\limits_{n\,=\,1}^{\infty }\frac{\lambda _{\alpha }%
}{n^{2}}\,\Gamma _{n}^{\prime }(\lambda _{\alpha })\right) \left\{ 1-\left( 
\frac{T_{\bot \alpha }}{T_{\Vert \alpha }}-1\right) \left( \frac{k_{\Vert
}^{2}v_{t_{\Vert \alpha }}^{2}}{2\omega ^{2}}\right) \right\}  \nonumber \\
&&  \nonumber \\
&&+2\sum_{\alpha }\frac{\omega _{p\alpha }^{2}}{\omega ^{2}}\lambda _{\alpha
}\,\Gamma _{0}^{\prime }(\lambda _{\alpha })\left\{ 1+\left( \frac{T_{\bot
\alpha }}{T_{\Vert \alpha }}-\,1\right) \left( \frac{k_{\Vert
}^{2}v_{t_{\Vert \alpha }}^{2}}{2\omega ^{2}}\right) \right\}.  \label{70}
\end{eqnarray}
Here we have expanded the plasma dispersion function$\,Z\left( \xi _{0\alpha
}\right) $ for $\left\vert \xi _{0\alpha }\right\vert \gg 1$.

To further simplify Eq.~\eqref{70}, we let $V_{A}^{2}\ll c^{2}$ and
take the small gyro radius limit (in which the argument of the Bessel
function is small), so that the term containing $%
\Gamma _{n}^{\prime }(\lambda _{\alpha })$ becomes negligible. This
leads to the result
\begin{equation}
\frac{c^{2}k^{2}}{\omega ^{2}}=1+\sum_{\alpha }\frac{\omega _{p\alpha }^{2}}{%
\Omega _{0\alpha }^{2}}-\sum_{\alpha }\frac{\omega _{p\alpha }^{2}}{\Omega
_{0\alpha }^{2}}\frac{k_{\bot }^{2}v_{\bot \alpha }^{2}}{\omega ^{2}}%
-\sum_{\alpha }\frac{\omega _{p\alpha }^{2}}{\Omega _{0\alpha }^{2}}\left( 
\frac{T_{\bot \alpha }}{T_{\Vert \alpha }}-1\right) \left( \frac{k_{\Vert
}^{2}v_{t_{\Vert \alpha }}^{2}}{2\omega ^{2}}\right),  \label{71}
\end{equation}
where we have used that $\Gamma _{0}^{\prime }(\lambda _{\alpha })=-1$ and
defined $\rho _{\alpha }^{2}\equiv v_{t\bot \alpha }^{2}/(2\Omega
  _{0\alpha }).$

For an electron-ion plasma, this dispersion relation can be written in
the form
\begin{equation}
\omega ^{2}=k^{2}V_{A}^{2}+k_{\bot }^{2}c_{s_{\bot }}^{2}+k_{\Vert
}^{2}c_{s_{\Vert }}^{2}\left\{ \frac{T_{\Vert i}}{T_{\Vert e}}\left( \frac{%
T_{\bot i}}{T_{\Vert i}}-1\right) +\left( \frac{T_{\bot e}}{T_{\Vert e}}%
-1\right) \right\}  \label{72}
\end{equation}%
where\[
\frac{\omega _{pi}^{2}}{\Omega _{i}^{2}}=\frac{c^{2}}{V_{A}^{2}}, \quad c_{s_{\bot }}^{2}=\frac{T_{\bot e}+T_{\bot i}}{m_{i}}, \quad \text{and} \quad c_{s_{\Vert }}^{2}=\frac{T_{\Vert e}}{m_{i}}.
\]

Equation~(\ref{72}), the modified dispersion relation for the fast
mode, incorporates the thermal anisotropy effects of the two
species. The parallel-propagating term displays an acoustic effect
associated with the parallel temperature, i.e., $c_{s_{\Vert }}^{2}$
which is sensitive to the thermal anisotropy. By contrast, the
acoustic effect $c_{s_{\bot }}^{2}$ associated with the
perpendicular-propagating term depends on the perpendicular
temperature, but receives no contribution from the temperature
anisotropy. In oblique propagation, the general fast mode has
therefore two distinct acoustic effects, one in the parallel and the
other in the perpendicular direction. The enhancement or the reduction
of the fast mode frequency depends upon the strength and signature of
the thermal anisotropies of the two species.

Figure~9 represents the general fast mode for small and for large
$\beta _{\Vert }$'s. The anisotropy effect becomes more prominent
when $\beta _{\Vert }$ is large.

Three instances of the general fast-mode dispersion
relation~(\ref{72}) deserve special mention:

\textbf{(i)} For parallel propagation (i.e., $k_{\bot }=0)$, Eq.~(\ref{72})
gives the Alfv\'{e}n-wave dispersion relation in Eq.~(\ref{40}), namely,
\begin{equation}
\omega ^{2}=k_{\Vert }^{2}V_{A}^{2}\left[ 1+\frac{\beta _{_{\Vert }}}{2}%
\left\{ \left( \frac{T_{\bot i}}{T_{\Vert }}-1\right) +\left( \frac{T_{\bot
e}}{T_{\Vert }}-1\right) \right\} \right].  \label{73}
\end{equation}
Here we have assumed that the two species have the same parallel temperature.

\textbf{(ii) }For perpendicular propagation ($k_{\Vert }=0$),
Eq.~(\ref{72}) yields the standard dispersion relation for the
magneto-sonic mode,
\[
\omega ^{2}=k_{\bot }^{2}\left\{ V_{A}^{2}+c_{s_{\bot }}^{2}\right\}.
\]

\textbf{(iii) }In the isotropic limit i.e., $T_{\bot \alpha }=T_{\Vert
\alpha }=T_{\alpha }$ Eq.~(\ref{72}) reduces to the standard fast mode
dispersion relation
\[
\omega ^{2}=k^{2}V_{A}^{2}+k_{\bot }^{2}c_{s}^{2},
\]
where 
\[
c_{s}^{2}=\frac{T_{e}+T_{i}}{m_{i}} 
\]\newline

\paragraph{Fast Mode Instability} $\quad$

$\quad$\\
Under the constraint 
\begin{eqnarray}
\frac{\left( 1+\dfrac{c_{s_{\bot }}^{2}}{V_{A}^{2}}\right) }{\left[ \dfrac{%
c_{s_{\Vert }}^{2}}{V_{A}^{2}}\left\{ \dfrac{T_{\Vert i}}{T_{\Vert e}}\left(
1-\dfrac{T_{\bot i}}{T_{\Vert i}}\right) +\left( 1-\dfrac{T_{\bot e}}{%
T_{\Vert e}}\right) \right\} -1\right] } &<&\frac{k_{\Vert }^{2}}{k_{\bot
}^{2}} \label{75},
\end{eqnarray}
the fast mode, Eq.~(\ref{72}), becomes unstable, since
\begin{eqnarray}
\omega =i\left[ k_{\Vert }^{2}V_{A}^{2}\left[ \frac{c_{s_{\Vert }}^{2}}{%
V_{A}^{2}}\left\{ \frac{T_{\Vert i}}{T_{\Vert e}}\left( 1-\frac{T_{\bot i}}{%
T_{\Vert i}}\right) +\left( 1-\frac{T_{\bot e}}{T_{\Vert e}}\right) \right\}
-1\right] -k_{\bot }^{2}V_{A}^{2}\left( 1+\frac{c_{s_{\bot }}^{2}}{V_{A}^{2}}%
\right) \right] ^{\frac{1}{2}}.\label{74}
\end{eqnarray}

The inequality~\eqref{75} requires that $\beta _{\Vert }=c_{s_{\Vert
  }}^{2}/V_{A}^{2}\gg 1$ and that $T_{\bot \alpha }<$\ $T_{\Vert
\alpha }$.

In Eq.~(\ref{74}), the parallel-propagating part represents the Alfv\'{e}n
wave with thermal anisotropy while the perpendicular part represents the
magneto-sonic wave, which tends to suppress instabilities. Therefore, fire-hose
instabilities may be damped for small perpendicular wavelengths, as
demonstrated by Fig.~10.\\

In the other limit, $\left\vert \xi _{n\alpha }\right\vert \leq 1$, we
expand the plasma dispersion function as
\begin{equation}
Z(\xi _{n\alpha })=i\sqrt{\pi }-2\xi _{n\alpha }\left( 1-\frac{2\xi
_{n\alpha }^{2}}{3}+\frac{4\xi _{na}^{4}}{15}+....\right) =i\sqrt{\pi }%
+\sum\limits_{l=0}^{\infty }\frac{(-2)^{l+1}}{\,\left( 2l+1\right) !!}\left( 
\frac{\omega -n\Omega _{\alpha }}{k_{\Vert }v_{t\Vert \alpha }}\right)
^{2l+1},  \label{76}
\end{equation}
and proceed to discuss the parallel-propagating whistler instability.\newline

\paragraph{Resonant Whistler Instability ( $\left\vert \xi _{n\alpha
}\right\vert \leq 1$) } $\quad$

$\quad$\\
With Eqs.~(\ref{10}), (\ref{11}),~and (\ref{13}), the dispersion relation
for R-waves in electron plasma becomes
\begin{eqnarray}
\frac{c^{2}k_{\parallel }^{2}}{\omega ^{2}} &=&1+\frac{\omega _{pe}^{2}}{%
\omega ^{2}}\left[ \left( \frac{T_{\bot e}}{T_{\Vert e}}-1\,\right) +\,\,i%
\sqrt{\pi }\frac{\omega }{k_{\Vert }v_{t_{\Vert e}}}\left\{ \frac{T_{\bot e}%
}{T_{\Vert e}}-\frac{\Omega _{oe}}{\omega }\left( \frac{T_{\bot e}}{T_{\Vert
e}}-1\right) \right\} \right],\label{77}
\end{eqnarray}
where we have kept only the leading term of the plasma dispersion
function for $\left\vert \xi _{n\alpha }\right\vert \leq 1.$

Under the subluminal condition, i.e., for $\omega \ll ck$ , the real and imaginary
parts of $\omega$ are
\begin{equation}
\Re\omega =\Omega _{0e}\left( 1-\frac{T_{\parallel e}}{T_{\perp e}}%
\right),  \label{78}
\end{equation}
and 
\begin{equation}
\Im\mathrm{\thinspace }\omega =\frac{k_{\Vert }v_{t_{\Vert e}}}{\sqrt{%
\pi }}\left( \frac{T_{\Vert e}}{T_{\bot e}}\right) \left[ \left( \frac{%
T_{\perp e}}{T_{\parallel e}}-1\right) -\frac{c^{2}k_{\Vert }^{2}}{\omega
_{pe}^{2}}\right].  \label{79}
\end{equation}
This defines the whistler instability, which occurs for waves
satisfying the wavenumber condition
\[
k_{\Vert }^{2}<\frac{\omega _{pe}^{2}}{c^{2}}\left( \frac{T_{\perp e}}{%
T_{\parallel e}}-1\right).
\]

The magnetic field generates real oscillations, but the growth rate
remains unaffected. These results coincide with those due to 
Lazar et al.,~\cite{lazar}. Figure~11 exhibits the effect of anisotropy on
the whistler instability. For the field-free case, i.e., $B_{0}=0$,
only the purely growing Weibel instability arises \cite{weibel}.

\section{\protect\bigskip Summary of Results and Discussion}
\label{sec:8}
On the basis of kinetic theory, we have extensively reviewed plasma
waves and instabilities. In particular, we find that for any
anisotropic equilibrium distribution, here included the
non-relativistic, relativistic, and ultra-relativistic magnetized
collisionless homogeneous plasmas, the electrostatic modes are
insensitive to thermal anisotropies, which affect only the
electromagnetic modes with magnetic field perturbations perpendicular
to the ambient magnetic field.

We have also derived expressions describing a general dielectric
tensor for a magnetized non-relativistic bi-Maxwellian plasma. In
alignment with the salient physical features of the problem, our
analysis showed that the dielectric tensor receives separate
contributions from the integrations over the perpendicular and the
parallel momentum components. The integration over the perpendicular
components leads to the functions $\Gamma _{n}(\lambda _{\alpha })$
and $\Gamma _{n}^{\prime }(\lambda _{\alpha })$, which are related to
the modified Bessel function$\,I_{n}(\lambda _{\,\alpha })$, while the
integration over the parallel components introduces the plasma
dispersion functions $Z\left( \xi _{n\alpha }\right) $ and $Z^{\prime
}\left( \xi _{n\alpha }\right)$.  From the resulting general
dielectric tensor we have obtained the dispersion relations for a
variety of modes and instabilities. 

We have also expanded the plasma dispersion functions in the limit
$\left\vert\xi_{n\alpha }\right\vert \gg 1$ to derive relatively
simple analytical expressions for the components of dielectric tensor
and to find the dispersion relations for those modes and
instabilities. For example, (i) for parallel propagation, we have
expanded the modified Bessel function as $k_{\bot }\rightarrow 0$ and
note that only $n=1$ terms survive. We have also derived the
dispersion relations for the R- and L- waves, whistler wave,
Alfv\'{e}n wave, Langmuir wave, non-resonant whistler instability,
Weibel instability and Alfv\'{e}n-wave instability; (ii) for
perpendicular propagation, we have let $k_{\Vert }\rightarrow 0$ and
noted that only the leading term ($l=0$) in the sum over $l$ survives
to derive the general dispersion relations for the X-, O-, and
Bernstein modes.  (iii) For oblique propagation, we have found the
general dispersion relations for the kinetic Alfv\'{e}n wave in
the kinetic and the inertial regimes and for the fast
mode. Additionally, we have discussed a few special cases of the fast
mode and the fast mode instability. For $\left\vert \xi _{n\alpha
  }\right\vert \leq 1$, we have derived the parallel-propagating
whistler instability.

In both limiting cases, as expected, the electrostatic modes are
unaffected by the thermal anisotropy. Among the electromagnetic modes,
the parallel-propagating modes, such as the R- and L- waves and the modes
derived from them (the whistler mode, pure Alfv\'{e}n mode, firehose
instability, and Weibel instability) are affected, while the
perpendicularly-propagating modes, such as the X-mode and the modes derived
from it (the pure transverse X- and Bernstein modes) are not.  The
O-mode, however, is sensitive to thermal anisotropies.

The thermal anisotropy affects the parallel
propagating modes via the acoustic effect, while it affects the perpendicular
propagating modes via Larmor-radius effects. For oblique propagation, both
effects appear, additively, in the Alfv\'{e}nic modes.

The effects of thermal anisotropies are more prominent in the kinetic
limit than the inertial limit. The parallel-propagating term of the
fast mode displays an acoustic effect with parallel temperature, i.e.,
$c_{s_{\Vert }}^{2}$ controlled by the thermal anisotropy. On the
other hand, the perpendicular-propagating term displays an acoustic effect
with perpendicular temperature, i.e., $c_{s_{\bot }}^{2}$, but it is
independent of the thermal anisotropy. In oblique propagation, the
general fast mode presents two distinct acoustic effects,
in the parallel and perpendicular directions.

For both the resonant and the non-resonant cases, the whistler and
Weibel instabilities can only exist in environments with $T_{\bot
  \alpha }>T_{\Vert \alpha }$ whereas the firehose and the general
fast-mode instabilities can only arise for $T_{\Vert \alpha }>
T_{\bot \alpha }.$

In conclusion, we find that the thermal anisotropy affects only the modes
with magnetic perturbations perpendicular to the ambient magnetic
field i.e., with $\mathbf{B}_{1}$ $\bot $ $\mathbf{B}_{0}$. The
anisotropy can either enhance or reduce the frequency domain of
waves and instabilities, depending on its strength and signature.

Our results may prove useful for studies of the thermally anisotropic
environments frequently found in astrophysical, space,  and
even laboratory plasmas.

\textbf{Acknowledgments}

We are thankful to the anonymous Referee for making several useful
suggestions to improve the quality of this review paper and to the
Office of the External Activities of the ICTP, Trieste, Italy, for
providing partial financial support to Salam Chair, at GC University
Lahore.

\textbf{Captions:}

Figure 1. This figure shows the geometrical representation of different modes. In all panels,the $z$ axis is aligned with the ambient field $\vec B_0$. The arrows indicate the directions of the electric field $\vec E_1$ , magnetic field $\vec B_1$ (blue bold arrow) and the wave vector $\vec k$ (red bold arrow). Panels (a) and (b) show an electromagnetic mode and an electrostatic mode propagating in the $z$ direction and. Panel (c) depicts an extraordinary mode, the coupling of an electrostatic and an electromagnetic modes, propagating in the $x$ direction. Panel (d) shows an ordinary mode propagating in the $x$ direction.

Figure 2.  This is the graphical illustration of circularly polarized electron R-wave. The normalized phase velocity  $\omega ^{2}/c^{2}k^{2}$ as function of normalized wave frequency $\omega /\Omega _{0e}$ which depicts the deviation from isotropic case with the change of thermal anisotropy values ( i.e., A=0(dotted), 5(green), 20(blue) , 40(red))  choosing the parameters  $\omega _{pe}/\Omega_{0e}=1.2,v_{t\Vert }/c=0.2$ \&\ $A=\left( T_{\bot e}/T_{\Vert e}\right) -1$. It shows that the phase velocity of R-wave increases with the increase in temperature anisotropy. 

Figure 3.   This is the graphical illustration of circularly polarized electron L-wave.  The normalized phase velocity  $\omega ^{2}/c^{2}k^{2}$ as function of normalized wave frequency $\omega /\Omega _{0e}$ which depicts the deviation from isotropic case with the change of thermal anisotropy values ( i.e., A=0(dotted), 5(green), 20(blue) , 40(red))  choosing the parameters  $\omega _{pe}/\Omega_{0e}=1.2,v_{t\Vert }/c=0.2$ \&\ $A=\left( T_{\bot e}/T_{\Vert e}\right) -1$. It shows that the phase velocity of L-wave increases with the increase in temperature anisotropy.

Figure 4.  This figure illustrate the graph of the normalized wave frequency  $\omega /\Omega _{e}$ of the whistler mode as a function of normalized wavenumber $ck/\omega _{pe}$ with the parameter  $\omega _{pe}/\Omega_{0e}=10,v_{t\Vert }/c=0.05$ \&\ $A=\left( T_{\bot e}/T_{\Vert e}\right) -1$ by choosing different values of  thermal anisotropy ( i.e., A=0(dotted), 5(green), 20(blue) , 40(red)). It shows that the wave frequency increases with the increase in temperature anisotropy.

Figure 5. The normalized growth rate  Im $\omega /\omega _{pe}$   of non-resonant whistler instability  ($\left\vert \xi_{n\alpha }\right\vert \gg 1$) as a function of normalized wavenumber $ck/\omega _{pe}$ with the parameter $\omega _{pe}/\Omega_{0e}=1.0,v_{t\Vert }/c=0.1$ \&\ $A=\left( T_{\bot e}/T_{\Vert e}\right) -1$  by choosing different values of  thermal anisotropy ( i.e., A= 0 (Black), 5 (green), 10 (red), 20 (Blue)).  By increasing the temperature anisotropy, The growth rate increases and threshold on wave number shifts towards the longer wavelength and thus enlarges the wave vector domain. This instability exist only in the plasma environments where with $T_{\bot \alpha }>T_{\Vert \alpha }$.

Figure 6. This figure depicts the behavior of X-mode for an electron plasma.
Panels (a), (c), and (e) show the behaviors for $n=1$ (dotted), and the
harmonics $n=2$ (green), and$n=3$ (red) , respectively, with $\lambda_e=0.05$, while
panels (b), (d), and (f) correspond to $n=1$, 2, and 3, respectively,
with $\lambda_e=8.0$.  The number of cutoffs and resonances increases
with $n$. The distinction between 
the resonance and the cutoff points is more prominent in the right
panels, i.e., for the large argument. The cutoff and
resonance points shift towards lower frequencies and hence expand the
propagation domain as $\lambda_e$ grows from 0.05 to 8.0.

Figure 7.  The electron Bernstein mode is illustrated graphically. The
dotted line represents the solution for $\lambda _{e}=1$ and the other
three curves represent $\lambda_{e}= 0.5$  (black), 1.0 (green),1.5 (blue), and 2.0 (red). The resonance points 
are fixed, but the cutoff points shift from right to left and reduce
the propagation domain as $\lambda _{e}$ grows. At higher harmonics
(not shown), the cutoff points become independent of $\lambda_{e}$.

Figure 8. The normalized frequency of  KAWs in kinetic limit  $\omega \left( k_{\Vert },k_{\bot }\right) /\omega _{pi}$ as a function of perpendicular and parallel normalized wavenumber i.e., $k_{\bot }\rho _{i}/\omega _{pi}),\beta _{\Vert }$ $%
=0.05$ $\ \ \ \&$ $\ A_{e}=A_{i}=A=\left( T_{\bot }/T_{\Vert }\right) -1$   by choosing different values of  thermal anisotropy ( i.e., A=0 (red), 10 (blue), 20(green) , 40(purple)). The frequency increases with the increase of temperature anisotropy value and the effect of temperature anisotropy is more prominent in high beta plasma environments  than the low beta one where  $T_{\bot \alpha }>T_{\Vert \alpha }$.

Figure 9. The normalized fast mode frequency $\omega \left( k_{\Vert },k_{\bot }\right) /\omega _{pi}$ as a funtion of perpendicular and parallel normalized wavenumber i.e.,  $(k_{\Vert}v_{A}/$ $\omega _{pi}$ \& $k_{\bot }v_{A}/\omega _{pi})$ with the parameters $\
A_{e}=A_{i}=A=\left( T_{\bot }/T_{\Vert }\right) -1;(a)\beta _{\Vert }$ $%
=0.05$ $\&$ ($b$)$\beta _{\Vert }$ $=10.0$   by choosing different values of  thermal anisotropy ( i.e., A=0 (red), 10 (blue), 15(green) , 30 purple). The frequency increases with the increase of temperature anisotropy value and the effect of temperature anisotropy is more prominent in high beta plasma environments  than the low beta one where  $T_{\bot \alpha }>T_{\Vert \alpha }$.

Figure 10.  The normalized growth rate  Im$\omega \left( k_{\Vert },k_{\bot }\right) /\omega _{pi}$  of fast mode instability as a function of perpendicular and parallel normalized wavenumber i.e.,  $(k_{\Vert}v_{A}/$ $\omega _{pi}$ \& $k_{\bot }v_{A}/\omega _{pi})$ with the parameters $A_{e}=A_{i}=A=\left( T_{\Vert }/T_{\bot }\right)\ \&$ $\beta _{\Vert }$ $=3.0$   by choosing different values of  thermal anisotropy ( i.e., A=1 (red), 2 (blue), 4 (green) , 9 (purple)). The growth rate enhances with the increase of temperature anisotropy but the perpendicular wave vector stabilizes this instability i.e., the magneto-sonic wave suppresses the fire-hose instability which exist only in the high beta plasma environment where with $T_{\bot \alpha }>T_{\Vert \alpha }$.

Figure 11. The normalized growth rate  Im $\omega /\omega _{pe}$   of non-resonant whistler instability  ($\left\vert \xi _{n\alpha
}\right\vert \leq 1$) as a function of normalized wavenumber $ck/\omega _{pe}$ with the parameters $v_{t\Vert }/c=0.1$ \&\ $A=\left( T_{\bot e}/T_{\Vert e}\right) -1$  by choosing different values of  thermal anisotropy i.e., A= 0 (Black), 5 (green), 10 (red), 20 (Blue).  By increasing the temperature anisotropy, The growth rate increases and threshold on wave number shifts towards the longer wavelength and thus enlarges the wave vector domain. This instability exist only in the plasma environment where with $T_{\bot\alpha }>T_{\Vert \alpha }$

\end{document}